\documentclass[aip,pra,preprint]{revtex4-1}

\usepackage{amssymb,amsmath,amsfonts,amstext,graphics,graphicx, subfigure}
\usepackage{epsfig}
\usepackage{dcolumn} 
\usepackage{bm}
\usepackage{relsize}
\usepackage{url} 
 \usepackage{latexsym}
 
\def\cald{\mathcal{D}}

\def\bq{\begin{equation}}
\def\eq{\end{equation}}
\def\bqy{\begin{eqnarray}}
\def\eqy{\end{eqnarray}}

\usepackage[colorlinks,citecolor=black,linkcolor=black]{hyperref}
\usepackage{color}

\def\al{\alpha}
\def\be{\beta}

\def\de{\delta}

\def\ep{\epsilon}

\def\ga{\gamma}

\def\la{\lambda}

\def\om{\omega}
\def\Om{\Omega}

\def\Ph{\Phi}

\def\si{\sigma}

\def\ta{\tau}

\def\p{\partial}

\begin{document}

\title{An  Action Principle for Relativistic Magnetohydrodynamics}
\author{Eric D'Avignon\footnote{cavell@physics.utexas.edu}}
\affiliation{Physics Department and Institute for Fusion Studies, The University of Texas at Austin, 
 Austin, TX 78712--1192}
\author{P. J. Morrison\footnote{morrison@physics.utexas.edu}}
\affiliation{Physics Department and Institute for Fusion Studies, The University of Texas at Austin, 
 Austin, TX 78712--1192}
 \author{F. Pegoraro\footnote{pegoraro@df.unipi.it}}
 \affiliation{Dipartimento di Fisica, Universit\`a di Pisa, 56127 Pisa, Italy}

\date{\today}

\begin{abstract}
A covariant action principle for ideal relativistic magnetohydrodynamics (MHD) in terms of natural Eulerian field variables is given.  This is done by generalizing the covariant Poisson bracket theory of Marsden et al.\  [Ann.\ Phys.\ {\bf169}, 29 (1986)], which uses a noncanonical bracket to effect constrained variations of an action functional.  Various implications and extensions of this action principle are also discussed.  Two significant by-products of this formalism are the introduction of a new divergence-free 4-vector variable for the magnetic field, and a new Lie-dragged form for the theory. 
\end{abstract}

\maketitle


\tableofcontents

\section{Introduction}
 
As a natural extension of early work on relativistic fluid mechanics \cite{synge}, Lichnerowicz and Anile developed a theory of relativistic magnetohydrodynamics \cite{lichnero, anile87,anile} paralleling the well-studied nonrelativistic version.  The primary assumption of MHD, that the fluid in question is charged but quasineutral, holds in  relativistic contexts of interest, although the definition of quasineutrality must be restated in terms of a 4-current.  As a result, relativistic MHD holds an important position in the field of relativistic computational modelling, with a variety of algorithms both suggested and implemented (e.g.\ Refs.~\onlinecite{komissarov,mignone,mignone2,zenitani,noble,gammie,cafe}).  The present paper explores the theoretical side of the subject, which recently has received less attention than the computational side.  In particular,  our  main contributions are i) the introduction of a new canonical 4-momentum, and a new divergenceless 4-vector to represent the magnetic field; ii) using the new variables to cast relativistic MHD into a covariant Poisson bracket formalism in terms of  Eulerian field variables; iii) investigating many properties of our new formalism, including several alternative brackets, a reformulation in differential-geometric concepts, and the consequences of a new gauge freedom.

Physicists know well the benefits of casting a theory into a  Hamiltonian or action principle mold, as our present work accomplishes.  In addition to being aesthetically appealing in its own right, this form has several practical advantages: (i) certain numerical algorithms  are based on such a structure (e.g.\  the recent works of 
Refs.~\onlinecite{zhou,kraus,shadwick}), while others can use said structure as a consistency check; (ii) finding the equations of motion in general coordinates, which Landau and Lifschitz called ``unsolved" for fluid mechanics, becomes straightforward; (iii) the formulation assists both the discovery and classification of constants of motion; (iv) a Hamiltonian structure provides a handy framework for equilibrium and stability analysis; (v) both Hamiltonian and action principle pictures provide a way of quantizing physical systems, tying into the field of ``quantum plasmas" currently receiving much attention.  The present work also necessitates a handful of new concepts (a modified enthalpy density, a momentum differing significantly from the standard kinetic momentum, and another ``momentum" conjugate to the magnetic field), which may provide new insight into this physical system.

It would be remiss for us not to mention previous attempts at providing an action principle for relativistic MHD.  Maugin \cite{maugin72} did provide a Lagrangian action principle, but in terms of Clebsch potentials, rather than the physical quantities themselves.  Alternatively,  Kawazura et al.\  \cite{KYF} have recently produced a useful Lagrangian variable action principle.  In future work we will show how these action principles relate to our own.  Meanwhile, the Poisson bracket structure of Morrison and Greene \cite{morgreene} was shown to be applicable to relativistic MHD  in terms of 3-vectorial quantities in a specific reference frame\cite{holm86}.  This  bracket  is  in effect a (3+1) split of the present  theory, which  uses only tensorial quantities  and does not require a choice of reference frame.  The chief advantage of the present work over Maugin's and Kawazura's is that it takes place in Eulerian variables, rather than Lagrangian ones: both the aforementioned theories require a  map to the physical Eulerian variables after the variational principle has been performed, adding an additional step that our formalism does not require.

The paper is organized as follows.  Section~\ref{sec:mhd} provides a review of MHD, starting with the nonrelativistic theory in Sec.~\ref{ssec:nrelmhd}  before describing  the  relativistic theory in  Sec.~\ref{ssec:relmhd}, where the new variable $h^\mu$ that describes the magnetic field  is introduced.  Section \ref{sec:action} then presents our new action principle using the new variable.    Here we first describe, in  Sec.~\ref{ssec:armhd}, the functional that serves as our action and show how conjugate variables arise from functional differentiation; then, in  Sec.~\ref{ssec:cpb}, we describe the covariant Poisson bracket formulation that provides our constrained variations.  Section \ref{sec:alternative} is dedicated to alternative brackets: first, in Sec.~\ref{ssec:const-bkt} we present a bracket with nontrivial Jacobi identity; in Sec.~\ref{ssec:bvp}, a bracket using a tensorial magnetic potential; in Sec.~\ref{ssec:3form}, the differential-geometric form of our main bracket, with several other quantities also presented in that form; finally, in Sec.~\ref{ssec:gravity}, we show how to couple our relativistic MHD theory to a fixed gravitational background.  In Sec.~\ref{sec:degeneracy} we discuss several features of our theory, including the nature of the divergence-free constraint and Casimirs.  Finally, in Sec.~\ref{sec:summary} we summarize our results.

\section{MHD Review}
\label{sec:mhd}

The equations of MHD, both nonrelativistic and relativistic, can be written in various ways in terms of different variables. In this section we gather together formulae  and well-known identities needed for the remainder of the paper.   The main new contribution of this section is the introduction of the variable $h^\mu$ of  \eqref{DefinitionOfh}.

\subsection{Nonrelativistic MHD -- two descriptions}
\label{ssec:nrelmhd}

First we give the equations of ideal nonrelativisitic ideal MHD, with the force law and Faraday's law expressed in two alternative ways:
\bqy
 \label{Neom}
\frac{\partial \mathbf{v}}{\partial t} &+& \left(\mathbf{v} \cdot \nabla \right)\mathbf{v} 
= -\frac{\nabla p}{\rho} + \frac{1}{4 \pi \rho}\big[\left(\nabla \times \mathbf{B}\right) \times \mathbf{B} \big]
 \\
& &\hspace{1.7 cm}  =-\frac{\nabla p}{\rho} + \frac{1}{4 \pi \rho} \nabla  \cdot \left(\mathcal{I}\,{B^2}/{2}- 
 \mathbf{B} \otimes \mathbf{B}\right)
  \label{NeomA}
\\
 \label{Nme}
\frac{\partial \mathbf{B}}{\partial t} &=& \nabla \times\left( \mathbf{v}\times\mathbf{B}\right) \\
& =& -\mathbf{B}\,  \nabla\cdot\mathbf{v}  +  \mathbf{B}\cdot\nabla \mathbf{v}
-  \mathbf{v}\cdot\nabla \mathbf{B}
 \label{NmeA}
 \\
 \frac{\partial \rho}{\partial t} &+& \nabla \cdot \left(\rho \mathbf{v}\right) = 0
\nonumber
\\
\frac{\partial s}{\partial t} &+& \mathbf{v} \cdot \nabla s = 0\,.
\nonumber
\eqy
Here $\rho$ is the fluid density, $p$ its pressure, $s$ its specific entropy, $\mathbf{v}$ the velocity field, and $\mathbf{B}$ the magnetic field.   In \eqref{NeomA} the symbol $\mathcal{I}$ represents the identity tensor.  The current $\mathbf{j}$ and electric field $\mathbf{E}$ have been eliminated from these equations, but they can be recovered from the ideal conductor Ohm's Law,  $\mathbf{E} + (\mathbf{v}/c)\times\mathbf{B} = 0$, and Amp\'{e}re's Law,  $\mathbf{j} = (c/4\pi)\nabla\times\mathbf{B}$.  

Observe the alternative versions of \eqref{Neom} and \eqref{Nme} given in  \eqref{NeomA} and \eqref{NmeA}, respectively.   These equations differ by terms involving  $\nabla\cdot\mathbf{B}$, and both Eqs.~\eqref{Nme} and \eqref{NmeA}  preserve the initial condition $\nabla\cdot\mathbf{B}=0$, which can be seen by rewriting \eqref{NmeA}:
\bq
\frac{\partial \mathbf{B}}{\partial t}  =  
  -\mathbf{B}\,  \nabla\cdot\mathbf{v}  +  \mathbf{B}\cdot\nabla \mathbf{v}
-  \mathbf{v}\cdot\nabla \mathbf{B} = \nabla \times\left( \mathbf{v}\times\mathbf{B}\right) - \mathbf{v}\,  \nabla\cdot\mathbf{B} \, .
\label{ConvertingBEquations}
\eq
Upon taking the divergence,
\bq
\frac{\partial \nabla\cdot\mathbf{B}}{\partial t} =   -\nabla\cdot( \mathbf{v}\,  \nabla\cdot\mathbf{B})\,.
\label{ptdivB}
\eq
Consequently, if $ \nabla\cdot\mathbf{B}$ is initially identically zero it remains so as well.   Equation \eqref{ConvertingBEquations} shows that forms \eqref{Nme} and \eqref{NmeA} are equivalent when the magnetic field is divergenceless, although the former reveals its Faraday law origin, while the latter shows an advected magnetic flux.  Geometrically \eqref{NmeA} is  $\p \mathbf{B}/\p t + \pounds_\mathbf{v} \mathbf{B}=0$, where $\pounds_\mathbf{v} \mathbf{B}$ is the Lie derivative of $\mathbf{B}$,  a vector density dual to a 2-form.    Similarly, Eqs.~\eqref{Neom} and \eqref{NeomA} differ by a $\nabla\cdot\mathbf{B}$ term, with the former revealing its Lorentz force origin via a clearly identified current, while the latter takes the form of a conservation law, which Godunov \cite{godunov} showed to be superior for numerical computation. 

We have distinguished these two forms because they possess different Hamiltonian structures.  In  Ref.~\onlinecite{morgreene} a Poisson bracket was given for the form with \eqref{Neom} and \eqref{Nme}, but this structure required building in the initial condition $\nabla\cdot\mathbf{B}=0$.   However, an alternative and more natural form was first given in 
Refs.~\onlinecite{morgreene82,morrison82}, which is entirely free from $\nabla\cdot\mathbf{B}=0$, it being only one possible choice for an initial condition.  Later in the paper we will demonstrate relativistic equivalents of both structures, and the two will also differ by the divergence of a 4-vectorial quantity; to be equivalent, said divergence must vanish, which will motivate our use of the new magnetic quantity $h^{\mu}$.

 Should one wish to add displacement current back into MHD, as is done in the most prevalent version of relativistic MHD, the momentum equation would have to be altered as follows:
\begin{equation} 
\label{NmhdD}
\frac{\partial \mathbf{v}}{\partial t} + \left(\mathbf{v} \cdot \nabla \right)\mathbf{v} = -\frac{\nabla p}{\rho} +
\frac{1}{4 \pi \rho}\left[\left(\nabla \times \mathbf{B} + 
\frac{\partial}{\partial t}\left(\frac{\mathbf{v}}{c^2}\times\mathbf{B}\right)\right) \times \mathbf{B} \right]\,.
\end{equation}
However, the new term, when compared to $\partial \mathbf{v} / \partial t$, scales as
\begin{equation*}
\frac{B^2}{4 \pi \rho c^2} = \left(\frac{v_A}{c}\right)^2\,,
\end{equation*}
where $v_A$ is the Alfv\'{e}n velocity.  In the nonrelativistic limit,  waves  involving disturbances of the matter must also travel much slower than the speed of light, allowing one to drop the displacement current.  This also means that relativistic MHD is free to add said displacement current back in (albeit constrained by Ohm's Law), while still reducing to conventional MHD in the nonrelativistic limit: one simply needs to keep in mind that said limit goes beyond just setting $v/c \rightarrow 0$.

\subsection{Relativistic MHD}
\label{ssec:relmhd}

Turning now to the description of relativistic MHD, we use signature and units such that 4-velocities have positive unit norms $u_{\mu}u^{\mu} = g_{\mu\nu}u^{\mu}u^{\nu}=1$, where the Minkowski metric $g_{\mu\nu}$ is given by $\mathrm{dia}(1,\!-1,\!-1,\!-1)$.  The 4-vector field $u^{\mu}$ will denote the plasma's 4-velocity at each point in spacetime; at each such point, this quantity will define a reference frame with locally vanishing 3-velocity, helpful for some purposes.  The fluid density is now $\rho = m n (1 + \epsilon)$, where $n$ is the baryon number density, $m$ is the fluid rest mass per baryon (including both proton and electron, for the typical case), and $\epsilon$ is the internal energy per baryon, normalized to $m$.  The specific entropy $s$ is unchanged, though later on it will prove more convenient to use the entropy density $\sigma = n s$.  We will suppose that the energy can be written $\epsilon(n,\si)$, hence $\rho(n,\si)$, in which case the pressure is given by 
\bq
p=n\frac{\partial \rho}{\partial n} + \sigma\frac{\partial \rho}{\partial \sigma} -\rho\,,
\label{pressure}
\eq
 which is just the first law of  thermodynamics, $Tds=d(\rho/n) + pd(1/n)$,  written in terms of $n$ and $\si$.  

In electromagnetism, having chosen a specific reference frame, one extracts the electric field 3-vector from the field tensor $F^{\mu\nu}$  by $\mathbf{E}^i= -F^{i0}$, $i=1,2,3$, while the magnetic field 3-vector $\mathbf{B}^i = \mathcal{F}_{i0}$,  where $\mathcal{F}_{\mu\nu}  = \epsilon_{\mu\nu\alpha\beta}F^{\alpha\beta}/2$ is the dual of $F^{\mu\nu}$.   Given $u^{\mu}$, one can also define the two 4-vectors $B^{\mu} \equiv \mathcal{F}^{\mu\nu}u_\nu=\gamma(\mathbf{v}\cdot\mathbf{B}, \mathbf{B} - \mathbf{v} \times \mathbf{E})$ and $E^\mu \equiv  {F}^{\mu\nu}u_\nu=  \gamma(\mathbf{v}\cdot\mathbf{E}, \mathbf{E} + \mathbf{v}\times\mathbf{B})$.  Note that $\mathbf{B}^i= B^i$ and $\mathbf{E}^i= E^i$ in the reference frame defined by $u^{\mu}$.  In terms of the  4-vectors $B^\mu$  and $E^\mu$ the field tensor has the decomposition
\begin{equation}
\label{FEB}
F^{\mu\nu} = \epsilon^{\mu\nu\lambda\sigma}B_{\lambda}u_{\sigma} + \left(u^{\mu}E^{\nu} - u^{\nu}E^{\mu}\right)\,,
\end{equation}
a form valid for any timelike 4-vector $u^{\mu}$.  One can also reverse this process by taking $B^{\mu}$ and $E^{\mu}$ to be fundamental, and then defining the field tensor $F^{\mu\nu}$ via \eqref{FEB}.  In this case, different values of $B^{\mu}$ and $E^{\mu}$ can give the same field tensor, for one can add any quantity proportional to $u^{\mu}$ to either 4-vector while leaving the field tensor unchanged; however, if the constraints $E^{\lambda}u_{\lambda} = B^{\lambda}u_{\lambda} = 0$ are imposed, then the representation is unique.  This multiplicity of representations of the field tensor will prove important later.

In MHD one  eliminates the electric field from the theory, if necessary using Ohm's Law to express it in terms of the fluid velocity and magnetic field.  In a relativistic context, this is done by setting $E^{\mu} = F^{\mu\lambda}u_{\lambda} = 0$, which gives $\mathbf{E} + \mathbf{v}\times\mathbf{B}=0$ (i.e. Ohm's Law) and, in a specific reference frame,
\bq
B^{\mu}={\gamma} \left(\mathbf{v}\cdot\mathbf{B},\, \frac{\mathbf{B}}{\ga^2}  + \mathbf{v} \, \left(\mathbf{v}\cdot\mathbf{B}\right)\right) \,.
\label{beebeebee}
\eq
For convenience  $b^{\mu} \equiv {B^\mu}/{\sqrt{4 \pi}}$ will be used,  in which case the MHD field tensor and its dual have the forms 
\bq
\label{FieldTensors} 
  F^{\mu\nu}= \sqrt{4\pi}\, \epsilon^{\mu\nu\lambda\sigma}b_{\lambda}u_{\sigma}   \quad \mathrm{and}\quad
\mathcal{F}^{\mu\nu}  = \sqrt{4\pi}\, (b^{\mu}u^{\nu} - u^{\mu}b^{\nu} )\,.
 \eq
 
Although \eqref{beebeebee} satisfies the restriction $b^{\lambda}u_{\lambda} = 0$, we noted earlier that this condition is not needed for a representation of the form of \eqref{FEB}.  One can, in fact, construct a family of vectors
\begin{equation}
\label{DefinitionOfh}
h^{\mu} = b^{\mu} + \alpha u^{\mu}
\end{equation}
where $\alpha$ is an arbitrary scalar field and now, in general, $h^\mu u_\mu = \alpha \neq 0$.  The field tensor $F^{\mu\nu}$ and its dual $\mathcal{F}^{\mu\nu}$ are unchanged when written in terms of $h^{\mu}$, i.e. 
\bqy
\label{hFieldTensors} 
  F^{\mu\nu}/\sqrt{4\pi}&=& \epsilon^{\mu\nu\lambda\sigma}b_{\lambda}u_{\sigma} = \epsilon^{\mu\nu\lambda\sigma}h_{\lambda}u_{\sigma}
  \\ \nonumber
\mathcal{F}^{\mu\nu}/ \sqrt{4\pi}  &=& b^{\mu}u^{\nu} - u^{\mu}b^{\nu} = h^{\mu}u^{\nu} - u^{\mu}h^{\nu}\,.
 \eqy
Because $b^{\mu}$ only appears in the equations of relativistic MHD via the form \eqref{FieldTensors}, one can just as easily use the quantity $h^{\mu}$, choosing $\alpha$ in order to give it some  useful property.    When constructing an Eulerian action principle (with covariant Poisson bracket) for relativistic MHD it will prove fruitful to do so.  The quantity $b_{\mu}b^{\mu}$, which appears in the stress-energy tensor and  will be seen in Sec.~\ref{sec:action} to appear in the action, evaluates to
\begin{equation}
b_{\mu}b^{\mu} = \frac{1}{4\pi}\left(E^2 - B^2\right) = 
-\frac{1}{4\pi}\left(\frac{\mathbf{B}\cdot\mathbf{B}}{\gamma^2} + \left(\mathbf{v}\cdot\mathbf{B}\right)^2\right)
=-\frac{1}{4\pi} B_{\mathrm{rest}}^2\,,
\nonumber
\end{equation}
where `rest' indicates a  rest frame quantity. 
Thus the 4-vector $b^{\mu}$ is spacelike.  However, since $h_{\mu}h^{\mu} = b_{\mu}b^{\mu} + \alpha^2$, the status of $h^{\mu}$ will depend on $\alpha$, remaining spacelike for small $\alpha$.

Each equation of relativistic MHD can be written as the vanishing of a divergence:
\begin{equation}\label{ContinuityEquation}
\p_\mu\!\left(n u^{\mu}\right) = 0
\end{equation} \begin{equation} \label{EntropyEquation}
\p_\mu\!\left(\sigma u^{\mu}\right) = 0
\end{equation} \begin{equation} \label{MaxwellEquation}
\p_\mu\, \mathcal{F}^{\mu\nu} = 0
\end{equation} \begin{equation} \label{MomentumEquation}
\p_\mu\, T^{\mu\nu} = 0\,. 
\end{equation}
Equations \eqref{ContinuityEquation} and \eqref{EntropyEquation} express conservation of particles and entropy, respectively.  
In addition,  \eqref{MaxwellEquation} provides the equivalent of the homogeneous Maxwell's equations; however,  one cannot call them Maxwell's equations without qualification, as the constraint $F^{\mu\nu}u_{\nu} = 0$ is already built in when one expresses $F^{\mu\nu}$ in terms of $b^{\mu}$ or $h^{\mu}$:
\begin{equation*}
\p_\nu\! \left(b^{\mu}u^{\nu} - u^{\mu}b^{\nu}\right)  = \p_\nu\! \left( h^{\mu}u^{\nu} - u^{\mu}h^{\nu} \right)  = 0\,.
\end{equation*}
This expression, of course, is the same whether $b^{\mu}$ or $h^{\mu}$ is used, as the quantity $\alpha$ cancels out.
Equation \eqref{MomentumEquation} gives conservation of stress-energy, where the stress-energy tensor $T^{\mu\nu}$ is considerably more complex when written in terms of $h^{\mu}$ rather than $b^{\mu}$:
\begin{equation} \label{StressEnergyTensor}
T^{\mu\nu} = T^{\mu\nu}_{fl} + T^{\mu\nu}_{EM}\,,
\end{equation}
where the fluid and field parts are
\bqy
T^{\mu\nu}_{fl} &=& \left(\rho + p\right)u^{\mu}u^{\nu} - p\, g^{\mu\nu}\,, \nonumber
\\
T^{\mu\nu}_{EM} &=& \frac{1}{4 \pi}\left(F^{\mu\lambda}F_{\lambda}^{\;\:\nu} + \frac{1}{4}g^{\mu\nu}F_{\lambda\sigma}F^{\lambda\sigma}\right)\nonumber \\
&=& - b^{\mu}b^{\nu} - \left(b_{\lambda}b^{\lambda}\right)u^{\mu}u^{\nu} + \frac{1}{2}g^{\mu\nu}b_{\lambda}b^{\lambda}
\label{Tb}\\
&=& - h^{\mu}h^{\nu} - \left(h_{\lambda}h^{\lambda}\right)u^{\mu}u^{\nu}
+ \left(h_{\lambda}u^{\lambda}\right)\left(h^{\mu}u^{\nu} + u^{\mu}h^{\nu}\right)
+ \frac{1}{2}g^{\mu\nu}\left(h_{\lambda}h^{\lambda} - \left(h_{\lambda}u^{\lambda}\right)^2\right) 
,
\label{Th}
\eqy
respectively.  Equation \eqref{Tb} is obtained by substitution of  the first of Eqs.~\eqref{FieldTensors}  and making use of the orthogonality condition $b^\la u_\la=0$, while \eqref{Th} follows from \eqref{hFieldTensors} without orthogonality.  We emphasize that, despite appearances, $T^{\mu\nu}_{EM}$ does not depend on one's choice of $\alpha$.  The field part $T^{\mu\nu}_{EM}$ depends on $b^{\mu}$ or $h^{\mu}$ only through the tensor $\mathcal{F}^{\mu\nu}$, in which, as previously noted, $\alpha$ cancels out.  Lastly, we note  it can be shown that this system preserves $b^\mu u_\mu=0$ and $u^\mu u_\mu=1$.  We next turn to the problem of devising an action principle for this system.

\section{Covariant Action Principle for  Relativistic MHD}
\label{sec:action}

 The covariant Poisson bracket formalism of Ref.~\onlinecite{MMMT} requires two parts: i) an action $S$ that is a covariant functional of the field variables and ii) a covariant Poisson bracket $\{\,,\,\}$ defined on functionals of the fields.   Instead of  the usual extremization $\delta S=0$, the theory arises from setting $\{F,S\}=0$ for all functionals $F$, which is in effect a constrained extremization.

A general Poisson bracket for fields $\Psi$ has the form
 \bq
 \{F, G\}=\int \! dz\,  \frac{\de F}{\de \Psi} \mathcal{J}\frac{\de G}{\de \Psi}\,, \nonumber
 \eq
where ${\de F}/{\de \Psi}$ is the functional derivative,  $dz$ is an appropriate spacetime measure, and $\mathcal{J}$ is a cosymplectic operator that provides  $\{F, G\}$ with the properties of antisymmetry and the Jacobi identity.  Thus 
\bq
\{F,S\}=0 \quad \forall \ F\quad  \Rightarrow \quad \mathcal{J} \frac{\de S}{\de \Psi}=0\,.
\label{variation}
\eq
If $\mathcal{J}$ is nondegenerate, i.e., has no null space, then \eqref{variation} is equivalent to ${\de S}/{\de \Psi}=0$ and the covariant Poisson bracket formalism reproduces the conventional variational principle. 
However, of interest here are matter models like MHD, which when written in terms of  Eulerian variables possesses nonstandard or noncanonical Poisson brackets (see e.g.\ Ref.~\onlinecite{morrison98}), for which $\mathcal{J}$ possess degeneracy that is reflected in the existence of so-called Casimirs (see Sec.~\ref{ssec:casimirs}).  For such systems the covariant Poisson bracket naturally enforces constraints.  In field theories that describe matter,  understanding the null space of  $\mathcal{J}$ may be a formidable  exercise\cite{yoshida14},  and finding nondegenerate coordinates, which are expected to  exist because of the Jacobi identity,  may only serve to obscure the structure of the theory.  

A variation that preserves the constraints, referred to as a dynamically accessible variation in Ref.~\onlinecite{pjmP89} (see also Ref.~\onlinecite{morrison98}), can be represented as
\bq
\de \Psi_{DA}= \{\Psi, G\}\,,
\label{DA}
\eq
for some functional $G$, whence
\bq
\de S=\int \! dz\, \frac{\de S}{\de \Psi} \, \de \Psi_{DA}
= \int \! dz\, \frac{\de S}{\de \Psi}\, \{\Psi, G\}
=  \{S, G\}=0\,,
\nonumber
\eq
which shows directly how the Poisson bracket effects the constraints without them being explicitly known.  

\subsection{Action and functional derivatives}
\label{ssec:armhd}

We construct our action $S$ in a straightforward fashion:
\bqy
S[n,\si, u, F] &=& \int\!d^4 x \, \left(
\frac{1}{2} \big(p + \rho \big)u_{\lambda}u^{\lambda} +  \frac{1}{2}\big( p -\rho\big)  - \frac{1}{16\pi}F_{\lambda\sigma}F^{\lambda\sigma}
\right)
\label{standard}\\
S[n,\si ,u, b] &=& \frac{1}{2}\int\!d^4 x \,\Big(\big(p+ \rho - b_{\lambda}b^{\lambda} \big)u_{\lambda}u^{\lambda} +   p -\rho \Big)
\label{MHDb}
\\
S[n,\si ,u, h] &=& \frac{1}{2} \int\!d^4 x \,\left( \big(p+ \rho  - h_{\sigma}h^{\sigma}\big)u_{\lambda}u^{\lambda} + 
 \big(h_{\lambda}u^{\lambda}\big)^2 +  p -\rho  \right)\,.
\label{RelativisticAction}
\eqy
Equation \eqref{standard} is the sum of the fluid action of Ref.~\onlinecite{MMMT}, where  thermodynamic variables $p$ and $\rho$ are considered to be functions of $n$ and $\sigma$, together with a standard expression  for  the electromagnetic action.  


In \eqref{MHDb} the MHD expression of \eqref{FieldTensors} has been substituted into $F_{\lambda\sigma}F^{\lambda\sigma}$ and finally in \eqref{RelativisticAction} we obtain our desired form in terms of $h^\mu$.   Observe that the integrand of \eqref{MHDb} when evaluated on the constraint $u_{\lambda}u^{\lambda} =1$ is the total pressure, fluid plus magnetic,  $p+  |b_{\lambda}b^{\lambda}|/2$.  This choice of action will be seen to give the desired field equations when inserted into the covariant Poisson bracket. 

From the action of \eqref{RelativisticAction} one derives a momentum $m_{\mu}$ by functional differentiation, 
\bq
m_{\mu} = \frac{\delta S}{\delta u^{\mu}} = \left(p+ \rho  - h_{\sigma}h^{\sigma}\right)u_{\mu} + \left(h_{\lambda}u^{\lambda}\right)h_{\mu} \equiv \mu u_{\mu} + \alpha h_{\mu}\,.
\label{MomentumDefinition}
\eq
The quantity 
\bq
\mu = p+ \rho  -h_{\lambda}h^{\lambda}
\label{muh}
\eq
 is a modified enthalpy density.  If $\alpha u^{\mu}$ is small compared to $b^{\mu}$, $h^{\mu}$ will be spacelike, leaving $\mu$ positive. 

Since $u^\mu$ and $b^\mu$ are independent of $\al$, expressions for them solely in terms of $m^\mu$ and  $h^\mu$ can be obtained.  Using $\alpha = h_{\lambda}u^{\lambda}$, which follows from  \eqref{DefinitionOfh}, and $u^{\mu} =  \left(m^{\mu} - \alpha h^{\mu}\right)/\mu$, which follows from \eqref{MomentumDefinition}, we have
\begin{equation*}
\alpha = h_{\lambda}u^{\lambda} = \frac{1}{\mu}\left(h_{\lambda}m^{\lambda} - \alpha h_{\lambda}h^{\lambda}\right)\,.
\end{equation*}  
Then, solving for $\alpha$ gives
\begin{equation}
\alpha = \frac{h_{\lambda}m^{\lambda}}{\mu + h_{\sigma}h^{\sigma}}\,.
\label{albe}
\end{equation}
Equation \eqref{albe}, incidentally, shows that $\alpha$ can be written entirely in terms of the field variables $m^\mu$ and $h^\mu$.  Thus, one can also write the variables $b^{\mu}$ and $u^{\mu}$ entirely in terms of the new ones:
\begin{equation} \begin{aligned} \label{VelocityAndField}
u^{\mu} &= \frac{m^{\mu}}{\mu} - \frac{h_{\lambda}m^{\lambda}}{\mu (\mu + h_{\sigma}h^{\sigma})}h^{\mu} \\
b^{\mu} &= h^{\mu}\left(1 + \frac{(h_{\lambda}m^{\lambda})^2}{\mu(\mu + h_{\sigma}h^{\sigma})^2}\right)
- \frac{h_{\lambda}m^{\lambda}}{\mu (\mu + h_{\sigma}h^{\sigma})}m^{\mu}\,.
\end{aligned} \end{equation}

Equations \eqref{VelocityAndField} are not invertible.  To see this consider a local frame in which $\mathbf{v}=0$, i.e., one where  $u^\mu=(1,\mathbf{0})$ and $b^\mu=(0,\mathbf{B})/\sqrt{4\pi}$.  In this frame
$h^\mu=(\al ,\mathbf{B}/\sqrt{4\pi})$ and $m_\mu=(p+\rho + B^2/4\pi, \al \mathbf{B}/\sqrt{4\pi})$.   Given any value of  $\al$ these equations are compatible with  \eqref{albe}, but produce the same rest frame values of  $b^\mu$ and $u^\mu$.  Thus, Eqs.~\eqref{VelocityAndField} are not one-one.   We will explore this degeneracy, which provides a kind of gauge condition, more fully in Sec.~\ref{sec:degeneracy}. 

Now we are in a position to obtain our action in terms of  the variables $m^\mu$ and $h^\mu$, which, due to the form of the upcoming bracket \eqref{RelativisticBracket}, are the appropriate variables for the action principle:
\begin{equation}
S[n,\si,m,h] =\frac12\int\!d^4 x \,  \left(\frac{m_{\lambda}m^{\lambda}}{\mu} - \frac{\left(h_{\lambda}m^{\lambda}\right)^2}{\mu(\mu+h_{\sigma}h^{\sigma})} + p-\rho  \right) \,.
\label{mhAction}
\end{equation}
Upon introducing  the ``mass'' matrix 
\bq 
\mathcal{M} \equiv \left( \begin{array}{cc}
\mu + \alpha^2  & \alpha  \\
\alpha & 1 
 \end{array} \right)\,, 
\label{massM}
\eq
 \eqref{mhAction} can be written compactly as
\bqy \nonumber
S &=&  \frac12\int\!d^4 x \,   \left(\, {\Psi}_\la \cdot \mathcal{M}^{-1}\!\cdot {\Psi}^\la  +  {h}_\lambda h^\lambda  - \alpha^2 + p - \rho \,
\right)\\ 
&=&  \frac12\int\!d^4 x \,   \left(u^\la m_\la  + b^\la h_\la +  {h}_\la h^\la  - \alpha^2 + p - \rho
\right)
\label{udotm}
\\ \nonumber
 &=&  \frac12\int\!d^4 x \,\left(\, {\Ph}_\la \cdot \mathcal{M}\!\cdot {\Ph}^\la + b_{\la}b^\la \, + p - \rho \right)
\\ \nonumber
 &=& \frac12\int\!d^4 x \,   \left(u^\la m_\la  + b^\la h_\la +  b_\la b^\la + p - \rho \right)
\eqy
where ${\Psi}^\la \equiv (m^\la,h^\la)$, $\Phi^\la \equiv (u^\la,b^\la)$ and $\cdot$ indicates summation over the $2\times 2$ matrix $\mathcal{M}$.  However, because the mass matrix \eqref{massM} depends  on the field variables  via $\mu$ and $\alpha$, as given by  \eqref{muh} and \eqref{albe}, the expression \eqref{mhAction} is superior for calculations; in addition, the mass matrix is inconsistent in units, so it would have to be normalized before, say, eigenvalue and eigenvector calculations could be done.  One possible normalization is given in \eqref{NormalizedMassMatrix} below.

After taking variations of the action, one may impose the constraint $u_{\lambda}u^{\lambda} = 1$.  In terms of the momentum $m^{\mu}$, this constraint becomes
\begin{equation} \label{MomentumSquared}
1 = u_{\lambda}u^{\lambda} = \frac{1}{\mu^2}\left(m_{\lambda}m^{\lambda} - 2\frac{\left(h_{\lambda}m^{\lambda}\right)^2}{\mu + h_{\sigma}h^{\sigma}} + \frac{\left(h_{\lambda}m^{\lambda}\right)^2}{\left(\mu + h_{\sigma}h^{\sigma}\right)^2}\left(h_{\tau}h^{\tau}\right)\right)\,.
\end{equation}
Thanks to the relations \eqref{VelocityAndField} and \eqref{MomentumSquared}, all functional derivatives of the action of \eqref{mhAction} can be reduced to simple expressions, provided \eqref{MomentumSquared} is applied only after functional differentiation.  To start with,
\bqy
\frac{\delta S}{\delta n} &=& \left(-\frac{m_{\lambda}m^{\lambda}}{2\mu^2} + \frac{\left(h_{\lambda}m^{\lambda}\right)^2}{2\mu^2(\mu + h_{\sigma}h^{\sigma})} + 
\frac{\left(h_{\lambda}m^{\lambda}\right)^2}{2\mu\left(\mu + h_{\sigma}h^{\sigma}\right)^2}\right)
\frac{\partial \mu}{\partial n} +\frac{1}{2}\frac{\partial p}{\partial n}- \frac{1}{2}\frac{\partial \rho}{\partial n}
\nonumber \\
&=& - \frac{\partial \rho}{\partial n}\,.
\label{Srho}
\eqy
Similarly,
\begin{equation}
\frac{\delta S}{\delta \sigma} = - \frac{\partial \rho}{\partial \sigma}\,.
\label{Ssig}
\end{equation}
The remaining functional derivatives are 
\bqy
\frac{\delta S}{\delta m^{\nu}} &=& \frac{m_{\nu}}{\mu} - \frac{(h_{\lambda}m^{\lambda})}{\mu(\mu+h_{\tau}h^{\tau})}h_{\nu} = u_{\nu}\,,
\label{Sm}
 \\ \nonumber
\frac{\delta S}{\delta h^{\nu}} &=&
 \frac{m_{\lambda}m^{\lambda}}{\mu^2}h_{\nu} - \frac{\left(h_{\lambda}m^{\lambda}\right)^2}{\mu^2(\mu + h_{\sigma}h^{\sigma})}h_{\nu} -
\frac{(h_{\lambda}m^{\lambda})}{\mu(\mu+h_{\sigma}h^{\sigma})}m_{\nu} \\
&=& \left(1 + 2\frac{\left(h_{\lambda}m^{\lambda}\right)^2}{\mu^2(\mu + h_{\sigma}h^{\sigma})} - \frac{\left(h_{\lambda}m^{\lambda}\right)^2}{\mu^2 \left(\mu + h_{\sigma}h^{\sigma}\right)^2}\left(h_{\tau}h^{\tau}\right)\right)h_{\nu}
\nonumber \\ \nonumber
 &&\hspace{2 cm} - \frac{\left(h_{\lambda}m^{\lambda}\right)^2}{\mu^2(\mu + h_{\sigma}h^{\sigma})}h_{\nu} -
\frac{(h_{\lambda}m^{\lambda})}{\mu(\mu+h_{\sigma}h^{\sigma})}m_{\nu} \\
&=& \left(1 + \frac{\left(h_{\lambda}m^{\lambda}\right)^2}{\mu(\mu+h_{\sigma}h^{\sigma})^2}\right)h_{\nu}
- \frac{(h_{\lambda}m^{\lambda})}{\mu(\mu+h_{\sigma}h^{\sigma})}m_{\nu}
\nonumber \\
&=& b_{\nu}\,.
\label{Sh}
\eqy
The compact result $\delta S/ \delta h^{\nu} = b_{\nu}$ gives a  meaning to $h^{\nu}$: it is a conjugate   to $b^{\nu}$, just as $m^{\nu}$ is to $u^{\nu}$.

\subsection{Covariant Poisson bracket and field equations}
\label{ssec:cpb}

The covariant Poisson bracket for relativistic MHD is obtained by  extending the nonrelativistic  bracket of  Refs.~\onlinecite{morgreene82,morrison82} to spacetime.  This is done by merely summing  over the four spacetime indices instead of the three spatial ones and altering a few signs.  However, a difficulty arises in choosing an appropriate equivalent of the nonrelativistic momentum and field, because the 4-vectorial equivalents of $\mathbf{M} = \rho \mathbf{v}$ and $\mathbf{B}$ will no longer produce the correct equations.  Instead, the 4-vectors $m^{\nu}$ and $h^{\nu}$ provide the appropriate replacements, giving the relativistic MHD bracket
\begin{equation} \label{RelativisticBracket} \begin{aligned}
\{F,G\} =& \int\!d^4 x \,  \bigg( n \left(\frac{\delta F}{\delta m_{\mu}}\partial_{\mu}\frac{\delta G}{\delta n} -
\frac{\delta G}{\delta m_{\mu}}\partial_{\mu}\frac{\delta F}{\delta n}\right)
+ \sigma \left(\frac{\delta F}{\delta m_{\mu}}\partial_{\mu}\frac{\delta G}{\delta \sigma} - 
\frac{\delta G}{\delta m_{\mu}}\partial_{\mu}\frac{\delta F}{\delta \sigma}\right) \\
& + m_{\nu}\left(\frac{\delta F}{\delta m_{\mu}}\partial_{\mu}\frac{\delta G}{\delta m_{\nu}} - 
\frac{\delta G}{\delta m_{\mu}}\partial_{\mu}\frac{\delta F}{\delta m_{\nu}}\right)
+ h^{\nu}\left(\frac{\delta F}{\delta m_{\mu}}\partial_{\mu}\frac{\delta G}{\delta h^{\nu}} - 
\frac{\delta G}{\delta m_{\mu}}\partial_{\mu}\frac{\delta F}{\delta h^{\nu}}\right) \\
& + h^{\mu}\left[\left(\partial_{\mu}\frac{\delta F}{\delta m_{\nu}}\right)\frac{\delta G}{\delta h^{\nu}} -
\left(\partial_{\mu}\frac{\delta G}{\delta m_{\nu}}\right)\frac{\delta F}{\delta h^{\nu}}\right]  
\Bigg)\,.
\end{aligned}
 \end{equation}
The bracket is complicated, but one can derive the equations of motion fairly quickly, thanks to the simple functional derivatives, as obtained in Eqs.~\eqref{Srho}, \eqref{Ssig}, \eqref{Sm}, and \eqref{Sh}, for the action of \eqref{mhAction}:
\begin{equation*}
\frac{\delta S}{\delta n} = -\frac{\partial \rho}{\partial n}\, ; 
\qquad \frac{\delta S}{\delta \sigma} = -\frac{\partial \rho}{\partial \sigma}\, ;
\qquad \frac{\delta S}{\delta m_{\nu}} = u^{\nu} \,;
\qquad \frac{\delta S}{\delta h_{\nu}} = b^{\nu} \,,
\end{equation*}
where $u^\mu$ and $b^\mu$ here are shorthands for their expressions in terms of the fields $m^\mu$ and $h^\mu$ as given by \eqref{VelocityAndField}. 

Using $F = \int\!d^4 x \,  n(x) \delta^4(x-x_0)$ in $\{F,S\} = 0$ gives, after an integration by parts,
\begin{equation*}
\p_\mu\!\left(n u^{\mu}\right)  = 0\,,
\end{equation*}
which is the continuity equation \eqref{ContinuityEquation}, evaluated implicitly at $x_0$; however, since that point is arbitrary, the result holds for the entire spacetime.  Going forward such niceties involving delta functions will be skimmed over.  In the same manner one also finds the adiabaticity equation \eqref{EntropyEquation} from a $\sigma$ variation.  

The $h^{\mu}$ variation gives
\begin{equation} \label{AlmostMaxwellEquation}
 \p_\nu\!\left(h^{\mu}u^{\nu}\right)  - h^{\nu} \p_\nu u^{\mu}  = 0\,.
\end{equation}
The above equations are not Maxwell's equations, although they are analogous to the nonrelativistic equation \eqref{NmeA}, since  they correspond to  $\pounds_u h^\mu=0$, the  Lie-dragging of the four-dimensional vector density $h^\mu$ by $u^\mu$.  The theory obtained from the variational principle can be viewed as a family of theories, only some of which correspond to physical systems.  However, if  $\p_\mu h^{\mu} = 0$, then one obtains the usual form of relativistic MHD.  The situation is exactly analogous to that in nonrelativistic Hamiltonian MHD, which can describe systems with $\nabla\cdot\mathbf{B} \neq 0$: in both cases, the physical systems are a subset of the full class of systems described by the formalism.    In the nonrelativistic case the condition $\nabla \cdot \mathbf{B} = 0$ is maintained by the dynamics and the similar situation that arises for $h^\mu$ will be shown in Sec.~\ref{ssec:gauge}.  There also exists an alternative bracket that builds in $\partial_\mu h^{\mu} = 0$, given later in Sec.~\ref{ssec:const-bkt}, where the constraint is enforced by the bracket's Jacobi identity.  In any event,  with $h^{\mu}$ thus specified,  we can subtract a term $u^{\mu}  \p_\nu h^{\nu}$ from \eqref{AlmostMaxwellEquation}, giving the usual equivalent of Maxwell's equations
\begin{equation*}
0 = \p_\mu\! \left(h^{\mu}u^{\nu} - u^{\mu}h^{\nu}\right) \,.
\end{equation*}

Finally, the $m^{\lambda}$ variation gives, after some work,
\bqy
0 &=& -n\partial^{\mu}\left(-\frac{\partial p}{\partial n}\right) - \sigma\partial^{\mu}\left(-\frac{\partial p}{\partial \sigma}\right) + m_{\nu}\partial^{\mu}\left(u^{\nu}\right) + \partial_{\nu}\left(m^{\mu}u^{\nu}\right)
\nonumber\\
&& \hspace{2 cm} + \ h_{\nu}\partial^{\mu}\left(b^{\nu}\right) - \partial_{\nu}\left(h^{\nu}b^{\mu}\right)
\nonumber\\
&=& -\partial^{\mu}p + \left(\mu u_{\nu} + \left(h_{\lambda}u^{\lambda}\right)h_{\nu}\right)\partial^{\mu}u^{\nu}
+ \partial_{\nu}\left(\mu u^{\mu}u^{\nu} + \left(h_{\lambda}u^{\lambda}\right)h^{\mu}u^{\nu}\right)
\nonumber\\
&& \hspace{2 cm} +\  h_{\nu}\partial^{\mu}\left(h^{\nu} - \left(h_{\lambda}u^{\lambda}\right)u^{\nu}\right)
- \partial_{\nu}\left(h^{\nu}h^{\mu} - \left(h_{\lambda}u^{\lambda}\right)h^{\nu}u^{\mu}\right)
\nonumber\\
&=&\p_\nu \bigg(\left(\rho + p - \left(h_{\lambda}h^{\lambda}\right)\right)u^{\mu}u^{\nu} 
+ g^{\mu\nu}\Big[-p + \frac{1}{2}\left(h_{\lambda}h^{\lambda} - \left(h_{\lambda}u^{\lambda}\right)^2\right)\Big]
\nonumber\\
&& \hspace{2 cm}-\  h^{\mu}h^{\nu} 
+ \left(h_{\lambda}u^{\lambda}\right)\left(h^{\mu}u^{\nu} + u^{\mu}h^{\nu}\right)\bigg)  \,,
\nonumber
\eqy
which is the momentum equation \eqref{MomentumEquation}.  Having been derived, it can be replaced with the much simpler, equivalent version involving $b^{\mu}$.

Now we have shown that the covariant Poisson bracket formalism produces the field equations of relativistic MHD.  In Secs.~\ref{ssec:gauge}  and \ref{ssec:gauge-set} we will probe more deeply the correspondence between the variables $(m^{\mu},h^{\mu})$ and $(u^{\mu},b^{\mu})$, exploring in particular how one might use the field equations in practice.  First, however, we will demonstrate several ways in which the bracket formalism can be modified.

\section{Alternative Brackets}
\label{sec:alternative}

In this section we present additional Poisson brackets.  The first (Sec.~\ref{ssec:const-bkt}) adds an extra constraint to \eqref{RelativisticBracket}, the second (Sec.~\ref{ssec:bvp}) rewrites the magnetic parts in terms of a tensor potential, the third (Sec.~\ref{ssec:3form}) recasts these terms in differential-geometric language, and the last (Sec.~\ref{ssec:gravity}) incorporates an arbitrary background gravitational field.   

\subsection{Constrained bracket}
\label{ssec:const-bkt}

Consider the magnetic field part of the bracket of
 \eqref{RelativisticBracket}, 
\bqy
\{F,G\}_h: &=& \int\!d^4 x \,\Bigg(  
h^{\nu}\left(
\frac{\delta F}{\delta m_{\mu}}\partial_{\mu}\frac{\delta G}{\delta h^{\nu}} - 
\frac{\delta G}{\delta m_{\mu}}\partial_{\mu}\frac{\delta F}{\delta h^{\nu}}\right)
\nonumber \\
&&\hspace{2 cm} +\  h^{\mu}\left[\left(\partial_{\mu}\frac{\delta F}{\delta m_{\nu}}\right)\frac{\delta G}{\delta h^{\nu}} -
\left(\partial_{\mu}\frac{\delta G}{\delta m_{\nu}}\right)\frac{\delta F}{\delta h^{\nu}}\right]  
\Bigg)\,. 
\label{hRelativisticBracket} 
\eqy
Just as the  nonrelativistic bracket of Ref.~\onlinecite{morgreene82,morrison82} has a counterpart in Ref.~\onlinecite{morgreene},
the terms \eqref{hRelativisticBracket} have an analogous relativistic counterpart that requires divergence-free magnetic fields, i.e. an $h^\mu$  such that $\p_\mu h^\mu =0$.  This relativistic  counterpart is simply given by an integration by parts of \eqref{hRelativisticBracket} and making use of $\p_\mu h^\mu =0$, i.e.,  
\bqy
\label{AlternativeRelativisticBracket} 
\{F,G\}_{\p h=0}: &=&
\int\!d^4 x \,\Bigg( 
h^{\nu}\left(\frac{\delta F}{\delta m_{\mu}}\partial_{\mu}\frac{\delta G}{\delta h^{\nu}} - 
\frac{\delta G}{\delta m_{\mu}}\partial_{\mu}\frac{\delta F}{\delta h^{\nu}}\right) \\ \nonumber
&&\hspace{2 cm}  + \ h^{\mu}\left[\left(\partial_{\mu}\frac{\delta F}{\delta h^{\nu}}\right)\frac{\delta G}{\delta m_{\nu}} -
\left(\partial_{\mu}\frac{\delta G}{\delta h^{\nu}}\right)\frac{\delta F}{\delta m_{\nu}}\right] 
\Bigg) \,.
\eqy 
The bracket is identical to \eqref{RelativisticBracket}, but for the swapped functional derivatives in the final line.  The action \eqref{RelativisticAction} is unchanged, as are the $n$ equation \eqref{ContinuityEquation} and the $\sigma$ equation \eqref{EntropyEquation}.  The $h^{\mu}$ gains an extra term, and may be written directly as the Maxwell-like equation
\begin{equation*}
\p_\nu \! \left(h^{\mu}u^{\nu} - u^{\mu} h^{\nu}\right) = \p_\nu  \mathcal{F}^{\mu\nu}  = 0
\end{equation*}
without yet imposing a condition on $h^{\mu}$.  Finally, the equation for $m^{\mu}$ ends up with a couple fewer terms than before, yielding
\begin{equation} 
\label{AlternativeMomentumEquation}
\p_\nu  T^{\mu\nu} + \big(h^{\mu} - \left(h^{\sigma}u_{\sigma}\right)u^{\mu}\big)
\p_\nu h^{\nu} = 0\,,
\end{equation}
where $T^{\mu\nu}$ is the (unchanged) stress-energy tensor \eqref{StressEnergyTensor}.

However, unlike the prior bracket \eqref{RelativisticBracket}, the bracket \eqref{AlternativeRelativisticBracket} fails to satisfy the Jacobi identity unless the condition $\p_\nu h^{\nu} = 0$ holds, as is shown in the Appendix.  On the plus side, the momentum equation \eqref{AlternativeMomentumEquation} is now reduced to its desired conservation form; on the minus side, the bracket is defined on a smaller class of functionals than our original bracket \eqref{RelativisticBracket}.  The original bracket always yields a momentum equation that is not only in conservation form, but also independent of $\alpha$; however, it will yield differing magnetic equations depending on $\alpha$, and only those corresponding to $\p_\nu h^{\nu}  = 0$ produce a Maxwell-like equation.  

We regard the first bracket \eqref{RelativisticBracket} to be superior, for then relativistic magnetohydrodynamics may be regarded as a specific example of a broader class of (mostly non-physical) dynamical systems, some of which may be of theoretical interest.  For instance, in the nonrelativistic case the broader class  have been argued to be superior for computational algorithms (see, e.g., Ref.~\onlinecite{godunov}), and although similar numerical techniques have been used for numerical relativity (e.g., Ref.~\onlinecite{komissarov}),  our formulation provides a fully covariant form analogous to nonrelativistic MHD that may provide advantages.  Moreover, they may correspond to exotic theories, such as those including magnetic monopoles.

\subsection{Bivector potential}
\label{ssec:bvp}

The divergence-free condition can be made manifest by introducing an antisymmetric bivector potential $A^{\nu \mu}$ such that
\bq
h^{\mu}=\p_{\nu} A^{\nu \mu}\,.
\label{bvp}
\eq
Such a representation gives rise to a gauge condition $A^{\mu \nu} \to   A^{\mu \nu} + \epsilon ^{\mu \nu\alpha\beta}\partial_\alpha \psi_\beta$, for chosen $ \psi_\beta$; such gauging  could be useful, but we will not explore this further here.

Assuming $F[h] = \bar{F}[A]$, i.e. functionals of the bivector potential obtain their dependence through $h$, we obtain
\bq
\de F=\int\!d^4x\,  \frac{\de F}{\de h^\mu}\de h^\mu= \int\!d^4x\,  \frac{\de \bar{F}}{\de A^{\nu\mu}}\de A^{\nu\mu} =\de \bar{F}\,.
\label{varAh}
\eq
Relate $\de h^{\mu}$ to $\de A^{\nu \mu}$ via \eqref{bvp} and insert $\de h^{\mu}= \p_{\nu} \de A^{\nu \mu}$ into the second equation of \eqref{varAh}.   Even assuming $\de A^{\nu\mu}$ is arbitrary, it only picks out the antisymmetric part of what it is contracted with, so we obtain the functional chain rule relation
\bq
 \frac{\de \bar{F}}{\de A^{\mu\nu}}= \frac12\left(\p_{\nu} \frac{\de F}{\de h^\mu}
 - \p_{\mu} \frac{\de F}{\de h^\nu}\right)\,.
 \label{AHchain}
\eq
Inserting \eqref{AHchain} into \eqref{AlternativeRelativisticBracket}  gives the compact expression
\bq
\{F,G\}_{A}: =
 2 \int\!d^4 x \, 
 \left( \p_{\al} A^{\al \nu} \right)  \left(\frac{\delta F}{\delta m_{\mu}}\ \frac{\delta G}{\delta A^{\nu\mu}} - 
\frac{\delta G}{\delta m_{\mu}}\ \frac{\delta F}{\delta A^{\nu\mu}}\right)\,.
\label{Abkt}
\eq
We will use this form in Sec.~\ref{ssec:casimirs}, where we discuss Casimir invariants. 

\subsection{3-Form bracket}
\label{ssec:3form}

For nonrelativistic MHD we observed in Sec.~\ref{ssec:nrelmhd}  that the magnetic equation may be written
$\p \mathbf{B}/\p t + \pounds_\mathbf{v} \mathbf{B}=0$, where $\pounds_\mathbf{v} \mathbf{B}$ is the Lie derivative of the   vector density $\mathbf{B}$ dual to a 2-form.  Thus one can write $B^i= \ep^{ijk} \om_{jk}$ and  $ \om_{jk}=B^i\ep_{ijk}/2$, where $i,j,k=1,2,3$. In terms of the 2-form the equation becomes $\p \om/\p t + \pounds_\mathbf{v} \om=0$, with $\pounds_\mathbf{v}$ now being the appropriate expression for the Lie derivative of a 2-form  in three dimensions (e.g., Ref.~\onlinecite{yano}).  In $n$ dimensions, an ($n\!\!-\!\!1$)-form has $n$ independent components.  This suggests we can introduce the dual 3-form  for relativistic MHD as follows:
\bq
\om_{\al\be\ga}= \ep_{\al\be\ga\de}\,h^\de  \qquad \mathrm{and}\qquad h^\de= \frac16\, \ep^{\al\be\ga\de} \om_{\al\be\ga}\,,
\label{3formDef} \nonumber
\eq
which shows that  $h^\mu$ is a vector density because it is the contraction of the tensorial three-form with $\ep_{\al\be\ga\de}$ a relative tensor of unit weight.   From the above it follows that the 3-form equation of motion is given by $\p \om/\p t + \pounds_u\om=0$.  If we denote by $F_m^\mu$ the 4-vector given by $\de F/\de m_\mu$, then the magnetic portion of the Poisson bracket in terms of the 3-form can be compactly written as follows:
\bq
 \{F,G\}_\om=\int \!\!d^4x\, \left(
 \frac{\delta F}{\delta \om_{\al\be\ga}}\left(\pounds_{G_m}\, \om\right)_{\al\be\ga} -  \frac{\delta G}{\delta \om_{\al\be\ga}}\left(\pounds_{F_m} \, \om\right)_{\al\be\ga}
 \right)\,.
 \label{3formbkt}
 \eq
Although similar expressions  in terms of Lie derivatives exist for all terms of all brackets, we are concentrating on the magnetic terms, which written out are
\bq \nonumber
\left(\pounds_{G_m}\, \om\right)_{\al\be\ga}= G_m^\mu \p_\mu\om_{\al\be\ga} 
+  \om_{\mu\be\ga} \p_\al G_m^\mu   + \om_{\al\mu\ga}   \p_\be  G_m^\mu  + \om_{\al\be\mu}   \p_\ga  G_m^\mu\,.
\eq
The transformation from the bracket $\{F,G\}_h$ of \eqref{hRelativisticBracket} to that of  \eqref{3formbkt} follows from a chain rule calculation similar to that described in Sec.~\ref{ssec:bvp}.   Thus, it satisfies the Jacobi identity because $\{F,G\}_h$ does, as shown directly in Appendix \ref{appendix}.

Relativistic MHD has a natural 3-form dual to $b^\mu$, viz.\ $F_{\la \si} u_\nu +F_{\si\nu} u_\la  +F_{\nu\la} u_\si$, which follows from the definition $b^\mu=\sqrt{4\pi} \ep^{\mu\nu\la\si} F_{\la \si} u_\nu/2$ with $u_\mu b^\mu=0$ and $F_{\mu\nu}u^\nu=0$.  The 3-form dual to $h^\mu$ can similarly be represented as $\om_{\la\si\nu}= \sqrt{4\pi}\left(F_{\la \si} w_\nu +F_{\si\nu} w_\la  +F_{\nu\la} w_\si\right)/6$, where $w_\mu \equiv (h^2u^\mu -\al h_\mu)/(b_\la b^\la)$ is designed so that $h^\mu w_\mu =0$ and $w_\mu u^\mu=1$ and evidently $\om_{\la\si\nu} h^\mu=0$.  Observe $w_\mu$ can be written in various ways using \eqref{VelocityAndField}, \eqref{albe},  and other expressions. 

The Jacobi identity for  the bracket with \eqref{3formbkt}  does not require closure of the 3-form.  However, if the 3-form $\om$ is exact then it can be written as the exterior derivative of a 2-form $\mathcal{A}_{\mu\nu}$ as follows:
\bq \nonumber
\om_{\al\be\ga}= \p_\al \mathcal{A}_{\be\ga} + \p_\be \mathcal{A}_{\ga\al} + \p_\ga \mathcal{A}_{\al\be}
\eq
and one can rewrite the bracket in terms of $A_{\mu\nu}$.  Instead of writing this out, we observe the bivector potential is given by 
\bq \nonumber
A^{\nu\mu} \equiv \frac12\, \ep^{\nu\mu\si\ta}  \mathcal{A}_{\si\ta}
\eq
and so the closed 3-form bracket is essentially given by \eqref{Abkt}. 

When the 3-form $\om_{\al\be\ga}$ is exact we have, for any 3-surface $\Om$ in our four-dimensional Minkowski space-time, Stokes' theorem
\bq
\int_\Om \, \om =\int_\Om \, \mathrm{d} \mathcal{A}= \int_{\p \Om} \mathcal{A}\,, 
\nonumber
\eq
where $\int_\Om \, \om$ contains the notion of `flux' in this setting.  If $\Om$ contains a time-like direction,  we can write this as a conservation law, but such $3+1$ splittings will not be considered here; instead, we refer to Ref.~\onlinecite{MMMT}. 
 
\subsection{Background gravity}
\label{ssec:gravity}

Now we generalize the full  formalism to curved spacetimes.  In this context, the equations \eqref{ContinuityEquation} - \eqref{MomentumEquation} are now written
\begin{equation}\label{CovariantContinuityEquation}
\left(n u^{\mu}\right)_{;\,\mu} = 0
\end{equation} \begin{equation} \label{CovariantEntropyEquation}
\left(\sigma u^{\mu}\right)_{;\,\mu} = 0
\end{equation} \begin{equation} \label{CovariantMaxwellEquation}
\mathcal{F}^{\mu\nu}_{\quad ; \, \nu} = 0
\end{equation} \begin{equation} \label{CovariantMomentumEquation}
T^{\mu\nu}_{\quad ;\,\nu} = 0\,,
\end{equation}
where the `;' denotes covariant derivative.

Three modifications to the previous action principle are required: (1) because all integrations have tensorial integrands, the integrations must take place over a proper volume $\sqrt{-g}\;d^4 x$; (2) $h^{\mu}$ should be treated as a contravariant vector, and $m_{\mu}$ as a covariant one, befitting their definitions (note that treating them any other way would introduce extra factors of $g^{\mu\nu}$ into the bracket); (3) functional derivatives should be defined in a way that makes them tensorial.  Specifically, for a field variable $v$, one implicitly defines the functional derivative via

\begin{equation*}
\left. \frac{d}{d \epsilon} \right|_{\epsilon = 0} F(v + \epsilon \delta v) = \int\!d^4 x \, 
 \frac{\delta F}{\delta v} \delta v\; \sqrt{-g} \,.
\end{equation*}

The action is now
\begin{equation*}
S = \frac12 \int\!d^4 x \, \left(\frac{g^{\lambda\sigma}m_{\lambda}m_{\sigma}}{\mu} - \frac{\left(h^{\lambda}m_{\lambda}\right)^2}{\mu(\mu+g_{\lambda\sigma}h^{\lambda}h^{\sigma})} + p-\rho \right)\sqrt{-g} 
\end{equation*}
and its functional derivatives are
\begin{equation*}
\frac{\delta S}{\delta n} = -\frac{\partial \rho}{\partial n}\,; \qquad
\frac{\delta S}{\delta \sigma} = -\frac{\partial \rho}{\partial \sigma}\,; \qquad
\frac{\delta S}{\delta m_{\mu}} = u^{\mu}\,; \qquad
\frac{\delta S}{\delta h^{\mu}} = g_{\mu\nu}b^{\nu}\,.
\end{equation*}

Finally, the bracket becomes
\begin{equation} \label{CovariantMHDBracket} \begin{aligned}
\{F,G\} =& \int\!d^4 x  \sqrt{-g}  \,  \Bigg( n \left(\frac{\delta F}{\delta m_{\mu}}\partial_{\mu}\frac{\delta G}{\delta n} -
\frac{\delta G}{\delta m_{\mu}}\partial_{\mu}\frac{\delta F}{\delta n}\right)
+ \sigma \left(\frac{\delta F}{\delta m_{\mu}}\partial_{\mu}\frac{\delta G}{\delta \sigma} - 
\frac{\delta G}{\delta m_{\mu}}\partial_{\mu}\frac{\delta F}{\delta \sigma}\right) 
 \\
& + m_{\nu}\left(\frac{\delta F}{\delta m_{\mu}}\partial_{\mu}\frac{\delta G}{\delta m_{\nu}} - 
\frac{\delta G}{\delta m_{\mu}}\partial_{\mu}\frac{\delta F}{\delta m_{\nu}}\right)
+ h^{\nu}\left(\frac{\delta F}{\delta m_{\mu}}\partial_{\mu}\frac{\delta G}{\delta h^{\nu}} - 
\frac{\delta G}{\delta m_{\mu}}\partial_{\mu}\frac{\delta F}{\delta h^{\nu}}\right) \\
&
 + h^{\mu}\left[\left(\partial_{\mu}\frac{\delta F}{\delta m_{\nu}}\right)\frac{\delta G}{\delta h^{\nu}} -
\left(\partial_{\mu}\frac{\delta G}{\delta m_{\nu}}\right)\frac{\delta F}{\delta h^{\nu}}\right] \Bigg)\,.
\end{aligned} \end{equation}
The $\partial_{\mu}$ operators inside the bracket are still just partial derivatives, but the presence of the metric will tend to convert them into covariant derivatives (see e.g.\ Ref.~\onlinecite{mtwgravitation},  Ch.~21).  After an integration by parts, the variation $\{F,S\} = 0$ of the test function $F = \int\!d^4 x \,  n(x) \;\delta^4(x - x_0) \;\sqrt{-g}$ gives
\begin{equation*}
\partial_{\mu}\left(n u^{\mu} \sqrt{-g}\right) = \sqrt{-g}\big(\p_\mu\left(n u^{\mu}\right) + 
n u^{\nu}\Gamma^{\mu}_{\ \nu \mu}\big) = \sqrt{-g} \left(n u^{\mu}\right)_{;\mu} = 0\,,
\end{equation*}
with a similar result obtaining for the $\sigma$ variation.  The $h^{\mu}$ variation once again requires special attention, as it gives
\begin{equation*}
\partial_{\nu}\left(h^{\mu}u^{\nu}\sqrt{-g}\right) - h^{\nu}\left(\partial_{\nu}u^{\mu}\right)\sqrt{-g} =
\sqrt{-g}\left(h^{\mu}u^{\nu}_{\ \nu} + h^{\mu}_{\ \nu}u^{\nu} - h^{\nu}u^{\mu}_{\ \nu}
+ h^{\mu}u^{\lambda}\Gamma^{\nu}_{\ \lambda\nu}\right) = 0\,.
\end{equation*}

This time we choose $\alpha$ so that $h^{\mu}_{\;\; ;\mu} = \p_\mu h^{\mu}  + h^{\nu}\Gamma^{\mu}_{\ \nu\mu} = 0$.  Similar considerations apply to this choice as in the special relativistic case.  Subtracting this expression and combining like terms then gives, with $\mathcal{F}^{\mu\nu} = h^{\mu}u^{\nu} - h^{\nu}u^{\mu}$, 
\begin{equation*}
\p_\nu \mathcal{F}^{\mu\nu}  + \mathcal{F}^{\mu\lambda}\Gamma^{\nu}_{\ \lambda\nu} + 
\mathcal{F}^{\nu\lambda}\Gamma^{\mu}_{\ \nu\lambda} = \mathcal{F}^{\mu\nu}_{\quad ;\nu} = 0\,.
\end{equation*}
Note that the third term is zero by the antisymmetry of $\mathcal{F}^{\mu\nu}$ and the symmetry of the covariant indices of $\Gamma^{\mu}_{\ \nu\lambda}$.

Finally, one obtains the momentum equation \eqref{CovariantMomentumEquation} by varying the test function $F = \int\!d^4 x \,  g^{\mu\nu}m_{\nu} \delta^4(x-x_0) \sqrt{-g}$.  This derivation is lengthy, and will only be summarized here: (1) the partial derivative terms appear, and combine, exactly as in the special-relativistic case; (2) the $T^{\mu\lambda}\Gamma^{\nu}_{\ \lambda\nu}$ terms come from taking the partial derivatives of $\sqrt{-g}$; (3) the $T^{\nu\lambda}\Gamma^{\mu}_{\ \nu\lambda}$ terms come from derivatives of extra factors of the metric $g^{\mu\nu}$, some of which come from its inclusion in the test function, others of which come from $\delta S/ \delta h^{\mu} = g_{\mu\nu}b^{\nu}$.

We conclude with an important note.  While we constructed the above formalism to handle curved spacetimes, it also applies to flat spacetimes with arbitrary coordinate systems, such as cylindrical, spherical, or toroidal coordinates.  The nonrelativistic version may be generalized the same way (altering volumes $d^3 x$ to proper volumes $\sqrt{g} d^3 x$), thus solving the problem of MHD coordinate changes in a pleasantly general way.  
%

\section{Degeneracy and Setting the Gauge}
\label{sec:degeneracy}

Now we consider various  issues pertaining to degeneracy.  In Sec.~\ref{ssec:casimirs} we obtain Casimir invariants, showing that the action $S$ is not unique.  Then in Sec.~\ref{ssec:gauge} we further explore the noninvertiblily of the  transformations from $(u^{\mu},b^{\mu})$ to $(m^{\mu},h^{\mu})$.  Finally, in Sec.~\ref{ssec:gauge-set} we discuss the how the  divergence-free condition on $h^\mu$ can be constructed for any problem. 

\subsection{Casimirs and degeneracy}
\label{ssec:casimirs}

As noted in Sec.~\ref{sec:action} the covariant Poisson bracket possesses degeneracy and associated Casimirs.  A  functional $C$ is a Casimir if it satisfies  
\bq
\{F,C\}=0 \qquad \forall   \ F\,.
\label{casimircon}
\eq   
Equation \eqref{casimircon}  should not be confused with the variational principle of \eqref{variation}, $\{F,S\}=0$ for all functionals $F$, for the former is an aspect of the bracket alone, and provides no equations of motion.  Because of the definition of $C$,  the action $S$ is not unique and can be replaced by $S+\lambda C$ for any Casimir C and any dimensionally appropriate number $\lambda$.  

Turning to the task of finding Casimirs, we use \eqref{casimircon} to provide functional equations for the Casimirs.  Although difficult to solve in general, some explicit solutions can be found, facilitated by our knowledge of Casimirs for nonrelativistic MHD\cite{morrison82,amp1}.  First, it  is easy to obtain a family of what we call the entropy Casimirs,
\bq
C_s=\int\!d^4x\, n f(\si/n)\,,
\eq
where $f$ is an arbitrary function. In the nonrelativistic case this is a generalization of the total entropy, for if $f=\si/n$ and $\si$ is the entropy per unit volume then $\int\!d^3x\, n f(\si/n)= \int\!d^3x\,  \si$ is the total integrated entropy. 

Next we seek a Casimir that is a relativistic version of the cross helicity $\int d^3x \, \mathbf{v}\cdot \mathbf{B}$.  For nonrelativistic MHD invariance of cross helicity requires a barotropic equation of state  and $\nabla\cdot \mathbf{B}=0$, so we  make analogous assumptions here.   We assume $\rho$ has no dependence on $\si$, and we implement the analogue of $\nabla\cdot \mathbf{B}=0$ by using the  antisymmetric bivector potential of Sec.~\ref{ssec:bvp},
$h^{\mu}=\p_{\gamma} A^{\ga \mu}$,  ensuring that $\p_{\mu}h^{\mu}=0$.  Using the bracket of \eqref{Abkt}
%
%
it is easy to show that the following generalization of the cross helicity is a Casimir:
\bq
C_{ch}=\int\!d^4x\,\frac{m_{\mu}}{n} \, \p_{\gamma} A^{\ga \mu}\, = \int \! d^4x \, \frac{m_{\mu}h^{\mu}}{n} \,.
\label{chCas}
\eq
This quantity ceases to be a Casimir when the divergence $h^{\mu}$ is nonzero. Observe that on the constraint $u_\la u^\la=1$, the integrand of \eqref{chCas} can be written as $m_{\mu}\, \p_{\gamma} A^{\ga \mu}/{n}=m_{\mu} h^\mu/{n}=\al (p + \rho)/n$, which follows from \eqref{albe}.  Since $\al$ does not exist in the original $(u^\mu,b^\mu)$ theory, this Casimir is a quantity  tied to the covariant bracket theory in terms of $(m^\mu,h^\mu)$. 

One also expects the existence of a magnetic  helicity Casimir, but the nature of linking in four dimensions makes the situation complicated.  Relativistic generalizations of magnetic helicity have been found in  Refs.~\onlinecite{pegoraro,KYF}, but we have yet to demonstrate that a quantity like either of these is in fact a Casimir.  We also anticipate the existence of additional Casimirs that are generalizations of the nonrelativistic ones found in 
Refs.~\onlinecite{padhye,padhye2}, but a full discussion of Casimirs  will await a future publication.  In any event, because of
the form $\Psi_{DA}$ as given by \eqref{DA}, we can be assured that the  extremization of our covariant bracket variation preserves any Casimirs that exist.

 
\subsection{Gauge degeneracy}
\label{ssec:gauge}

In Sec.~\ref{ssec:armhd} we noted that Eqs.~\eqref{VelocityAndField} are not invertible.   This lack of invertibility, which arises from the gauge freedom associated with $\al$, can be understood in  greater generality.   

Because the degeneracy is not associated with the thermodynamic variables $\rho$ and $p$, we move them out  by introducing the following scaled variables:
\begin{equation*}
h= (\sqrt{p +\rho}) \, \bar{h}\,,\quad
m= ({p +\rho}) \, \bar{m}\,,\quad
b= (\sqrt{p +\rho}) \, \bar{b}\,,\quad
u=\bar{u}\,, \quad
\alpha = (\sqrt{p + \rho}) \, \bar{\alpha} \, ,
\end{equation*}
In terms of these variables \eqref{VelocityAndField} becomes 
\bq
\bar{\Phi}= \bar{\mathcal{M}}^{-1}\!\cdot \bar{\Psi}\,,
\label{trans}
\eq
with
\bq
\bar{\mathcal{M}}^{-1}=\frac1{\bar{\mu}}\left[
\begin{array}{cc}
1 & - \bar{\al}\\ 
-\bar{\al} & \quad \ \bar{\mu} +\bar{\al}^2
\end{array}
\right]\,,
\qquad  \bar{\mathcal{M}} = \left[
\begin{array}{cc}
\bar{\mu} +\bar{\al}^2 &\quad \  \bar{\al}\\ 
\bar{\al} & 1
\end{array}
\right]\,,
\label{NormalizedMassMatrix}
\eq
and $\bar\Phi=(\bar u, \bar b)$ ,  $\bar\Psi=(\bar m, \bar h)$.  The quantity $\bar{\mu} \equiv {1-\bar{h}^2}$ is a normalized $\mu$, and the quantity $\bar\alpha$ satisfies $\bar{\al} = \bar{m}_{\nu}\bar{h}^{\nu}=\bar{u}_{\nu}\bar{h}^{\nu}$.  Varying \eqref{trans} gives
\begin{equation*}
\de \bar{\Phi}= \bar{\mathcal{M}}^{-1}\!\cdot \de\bar{\Psi} + \frac{\p \bar{\mathcal{M}}^{-1}}{\p \bar{\al}}\cdot \bar{\Psi} \, \de \bar{\al}
+ \frac{\p \bar{\mathcal{M}}^{-1}}{\p {\bar{\mu}}}\cdot \bar{\Psi} \, \de \bar{\mu}\,.
\end{equation*}
Degeneracy follows if we can find a nonzero $\de \bar{\Psi}$ giving  $\de \bar{\Phi}=0$.  Such would be given by 
\bqy
\de\bar{\Psi} &=&- \bar{\mathcal{M}} \cdot \frac{\p \bar{\mathcal{M}}^{-1}}{\p \bar{\al}}\cdot \bar{\Psi} \, \de \bar{\al}
-\bar{\mathcal{M}}\cdot \frac{\p \bar{\mathcal{M}}^{-1}}{\p {\bar{\mu}}}\cdot \bar{\Psi} \, \de \bar{\mu}
\nonumber\\
&=&- \bar{\mathcal{M}} \cdot \frac{\p \bar{\mathcal{M}}^{-1}}{\p \bar{\al}}\cdot \mathcal{M} \cdot \bar{\Phi} \, \de \bar{\al}
-\bar{\mathcal{M}}\cdot \frac{\p \bar{\mathcal{M}}^{-1}}{\p {\bar{\mu}}}\cdot  \bar{\mathcal{M}} \cdot \bar{\Phi}\, \de \bar{\mu}
\nonumber\\
 &=& \frac{\p \bar{\mathcal{M}}}{\p \bar{\al}}\cdot \bar{\Phi} \, \de \bar{\al} +
 \frac{\p \bar{\mathcal{M}}}{\p {\bar{\mu}}}\cdot \bar{\Phi} \, \de \bar{\mu}
 \nonumber
\\
 &=&{\de \bar{\al}} \left[
\begin{array}{cc}
 2\bar{\al} &\ \ 1\\ 
1 & \ \ 0 
\end{array}
\right]\cdot\bar{\Phi}
+ {\de \bar{\mu}}   \left[
\begin{array}{cc}
1 &\ \   0\\ 
0 & \ \   0
\end{array}
\right]\cdot\bar{\Phi}\,.
\label{deub}
\eqy
Thus from \eqref{deub},  $\de \bar{m}^{\nu} = (2\bar{\al} \bar{u}^{\nu} + \bar{b}^{\nu}) \de \bar{\al} + \bar{u}^{\nu}  \delta \bar{\mu}$ and  
$\de \bar{h}^{\nu} = \bar{u}^{\nu} \, \de \bar{\al}$.   Using $\de\bar{\mu}=-2\bar{h}^{\nu} \de \bar{h}_{\nu} =-2 \bar{h}^{\nu} \bar{u}_{\nu} \, \de \bar{\al}= -2  \bar{\al}\,  \de \bar{\al} $,  the two conditions imposed by \eqref{deub} are
\bq
\de \bar{h}^{\nu} = \bar{u}^{\nu} \, \delta \bar{\alpha} \qquad \mathrm{and}\qquad 
\de \bar{m}^{\nu} =   \bar{b}^{\nu} \, \de \bar{\alpha}  \, ,
\label{symvar}
\eq
reiterating our earlier point that $\alpha$ can vary while leaving $u^{\mu}$ and $b^{\mu}$ unchanged.

In terms of the scaled variables the action becomes
\begin{equation}
S[n,\si,\bar{m},\bar{h}] =\frac12\int\!d^4 x \, \left( \frac{p + \rho}{\bar{\mu}}
\left(\bar{m}_{\lambda}\bar{m}^{\lambda} -  \left(\bar{h}_{\lambda}\bar{m}^{\lambda}\right)^2\right) + p-\rho
  \right) \,.
\label{mhscaledAction}
\end{equation}
Now if we consider variation of the integrand of \eqref{mhscaledAction} with variations given by \eqref{symvar}, and restrict to the constraint  $u_\mu u^\mu =1$ as given by the scaled version of  \eqref{MomentumSquared}, then the action is easily seen to be invariant.   Using the scaled action in the form of \eqref{udotm},  the integrand becomes upon variation $(p +\rho)(\bar{u}_\la \de \bar{m}^\la +\bar{b}_\la \de \bar{h}^\la) + \bar{h}_\la \de \bar{h}^\la - \bar{\al} \de \bar{\al}$, which vanishes upon insertion of \eqref{symvar}, with the first two terms vanishing individually because   $\bar{u}_\la \bar{b}^\la=0$.   Thus,  degeneracy appears as one transitions from \eqref{MHDb} to \eqref{RelativisticAction}.   We add that in scaled variables $\mathcal{F}\sim \bar{u}^\mu \bar{b}^\nu -  \bar{b}^\mu \bar{u}^\nu \sim  \bar{u}^\mu \bar{h}^\nu -  \bar{h}^\mu \bar{u}^\nu$; thus, at fixed $\bar{u}^\mu$, $\de\mathcal{F}\sim   \bar{u}^\mu \de \bar{h}^\nu - \de  \bar{h}^\mu \bar{u}^\nu= 0$.  

However, we also require that the equations of motion in terms of $(u^{\mu},b^{\mu})$ stay unaffected by the degeneracy in $(m^{\mu},h^{\mu})$.  This requires $\partial_{\mu} h^{\mu} = 0$, as we earlier discussed in the context of the magnetic equation \eqref{AlmostMaxwellEquation}.  Written in full, this condition becomes
\begin{equation} \label{ConditionOnAlpha}
\partial_{\mu} \left(\alpha u^{\mu}\right) = -\partial_{\mu} b^{\mu}\,.
\end{equation}
As usual in the case of extra degeneracy, the system now possesses an additional symmetry, for one can add to $\alpha$ any solution $\Delta \alpha$ of the continuity equation $(\Delta \alpha u^{\mu})_{,\mu} = 0$ while leaving the dynamics unchanged.  This is not as powerful as choosing $\alpha$ freely, as \eqref{symvar} seemed to imply, but we will show in Sec.~\ref{ssec:gauge-set} that it is nearly as powerful.  We hope to further explore the consequences of this new symmetry in future work.

Our system's degeneracy is related to  the  adaptation of Goldstone's theorem\cite{nambu-JL,nambu,nambu2,goldstone}  described  in  Ref.~\onlinecite{pjmE86},  where it  was proven in the context of degenerate Poisson brackets with Casimir invariants that  nonrelativistic Alfv\`en waves   associated with degeneracy can be thought of as an analog of Goldstone modes.  A similar interpretation arises here in this covariant relativistic MHD setting, but discussion is beyond the scope of the present work. 
 
\subsection{Setting the gauge}
\label{ssec:gauge-set}

Given a relativistic MHD problem posed in terms of $(u^\mu,b^\mu)$,  we must determine the associated problem in terms of $(m^\mu,h^\mu)$, and this requires the determination of $\al$, which amounts to setting the gauge so that 
$\p_\mu h^\mu =0$.  Doing so may seem difficult on first sight, but in fact turns out to be simple.  Since this idea sits at the crux of our formalism, we will explain it is some detail. 

Posing a relativistic MHD problem requires one specify $(u^\mu,b^\mu)$ as well as $n$ and $\si$  on a space-like 3-volume, $\Om\subset \cald$, where is $\cald$ is our four-dimensional space-time.  In addition, a physical problem will have  initial conditions that satisfy $u_\la u^\la=1$ and $u_\la b^\la=0$.  Using $u^\al \p_\al = \p /\p \tau$ where $\tau$ is the proper time measured by an observer comoving with a flow line, one can choose $\tau=0$ to correspond to  the state specified on $\Om$  and then propagate values off of $\Om$ by using the equations of motion to determine $\p b^\mu/\p \tau$,  $\p u^\mu/\p \tau$, $\p n/\p \tau$, and $\p \si/\p \tau$ at $\tau=0$.  This is the standard scenario for a Cauchy problem, and many references for both MHD and relativistic fluids  (e.g., Refs.~\onlinecite{lichnero,anile})  describe  this in  detail.   One can imagine an exotic flow in which there exist spacetime points not connected to $\Omega$ by any flow lines; however, a modest boundedness condition excludes such cases.  

The present situation is complicated by the fact that given $b^\mu$ on $\Om$ at $\tau=0$ we  must also have that  $\p_\mu h^\mu=0$ for all time,  in order for our  $(m^\mu,h^\mu)$ dynamics to coincide with the physical $(u^\mu,b^\mu)$ dynamics.   Fortunately, $\p_\mu h^\mu=0$ is maintained in time if it is initially true on $\Omega$.  To see this we act on  \eqref{AlmostMaxwellEquation} with  
$\partial_{\mu}$ and obtain  
$
\partial_{\nu}(u^{\nu} \partial_\mu h^{\mu}) 
= u^{\nu}  \partial_{\nu}(\partial_\mu h^{\mu}) + (\partial_{\nu}u^{\nu})(\partial_\mu h^{\mu})  
=0 
$ or 
\bq
\frac{\p (\partial_\mu h^{\mu})}{\p \tau} + (\partial_{\nu}u^{\nu})(\partial_\mu h^{\mu})  
=0 \,,
\label{Dh}
\eq
an equation analogous to \eqref{ptdivB} for nonrelativistic MHD.  
From \eqref{Dh}, one concludes that if  $\partial_\mu h^{\mu}=0$ on $\Om$ at $\tau=0$,  then $\partial_\mu h^{\mu}$ remains zero for all time.  Thus,  one can solve the $(m_{\mu},h_{\mu})$ equations and uniquely obtain the $(u_{\mu},b_{\mu})$ via \eqref{VelocityAndField} --  provided one can `set the gauge', i.e., find an $\al$ such that $\partial_\mu h^{\mu}=0$ on $\Om$ at $\tau=0$ consistent with the $(u^\mu,b^\mu,n,\si)$ of our posed problem. 

We will first consider a special example of setting the gauge, corresponding to the case described in  Sec.~\ref{ssec:armhd}.  We are given the MHD problem with initial conditions $\mathbf{v}(0,\mathbf{x})\equiv 0$, i.e., $u^\mu(0,\mathbf{x})=(1,\mathbf{0})$ and $b^\mu(0,\mathbf{x})=(0,\mathbf{B}(0,\mathbf{x}))/\sqrt{4\pi}$ on the space-like 3-volume $\Om$ with coordinates $\mathbf{x}$, and we wish to obtain an $h^\mu(0,\mathbf{x})=(\al ,\mathbf{B}/\sqrt{4\pi})$ and $m_\mu(0,\mathbf{x})=(p+\rho + B^2/4\pi, \al \mathbf{B}/\sqrt{4\pi})$ such that $\p_\mu h^\mu(0,\mathbf{x})=0$.  Denoting $\p_0 \al = \al_t$, etc.,   gives the condition
\bq
0=\p_\mu h^\mu(0,\mathbf{x})=\frac{1}{\sqrt{4\pi}}\left(\ga_t \mathbf{v}\cdot \mathbf{B} + \ga \mathbf{v}_t\cdot \mathbf{B}
+   \ga \mathbf{v}\cdot \mathbf{B}_t +  \al_t \sqrt{4\pi}\right)  +\nabla\cdot \mathbf{h}\,,
\label{ph}
\eq
where $\mathbf{h}$ is the spatial part of $h^\mu$.  Evaluating \eqref{ph}   on the initial condition gives
\bq \nonumber
0=\mathbf{v}_t\cdot \mathbf{B}(0,\mathbf{x}) +  \al_t(0,\mathbf{x}) \sqrt{4\pi}  
+  \nabla\cdot\mathbf{B}(0,\mathbf{x})\,,
\eq
whence, with $\nabla\cdot\mathbf{B}(0,\mathbf{x})=0 $,   we conclude that 
\bq
0=\mathbf{v}_t\cdot \mathbf{B}(0,\mathbf{x}) +  \al_t(0,\mathbf{x}) \sqrt{4\pi}=-\frac1{\rho}\nabla p\cdot \mathbf{B}(0,\mathbf{x}) +  \al_t (0,\mathbf{x})\sqrt{4\pi} \,,
\nonumber
\eq 
using the MHD momentum equation in the last step.  Thus $\al_t(0,\mathbf{x})= (\sqrt{4\pi}{\rho})^{-1}\nabla p\cdot \mathbf{B}(0,\mathbf{x})$ on $\Omega$ will assure $\p_\mu h^\mu=0$ for all time.  Observe,  $\al(0,\mathbf{x})$ has not been specified -- we are free to choose it as we please.  In doing so we will obtain different initial conditions $m^\mu(0,\mathbf{x})$ and $h^\mu(0,\mathbf{x})$ and these can be chosen for convenience.  Finally, if we solve our equations for $m^\mu$ and $h^\mu$ and obtain their  values at any later time, insert them into \eqref{VelocityAndField},  then values of $u^\mu$ and $b^\mu$ thus obtained are solutions of the relativistic MHD equations.

Now let us consider the general case, beginning with the expression
\bq
\p_\mu h^\mu= \p_\mu b^\mu + \p_\mu(\al u^\mu)= \p_\mu b^\mu + \p_\mu(\al n u^\mu/n)
= \p_\mu b^\mu + n \,\frac{\p }{\p \tau}\left(\frac{\al}{n}\right)\,,
\label{pb1}
\eq
where the last equality follows from \eqref{ContinuityEquation}.  Upon contracting $\p_\nu\! \left(b^{\mu}u^{\nu} - u^{\mu}b^{\nu}\right)=0$ with $u_\mu$ we obtain
\bq
\p_\nu b^\nu= u_\nu \frac{\p b^\nu}{\p \tau}= - b^\nu \frac{\p u_\nu}{\p \tau}\,.
\label{pb2}
\eq
Consequently, \eqref{pb1} and  \eqref{pb2} imply
\bq
\frac{\partial}{\partial \tau}\left(\frac{\alpha}{n}\right) = \frac{b^{\nu}}{n}\frac{\partial u_{\nu}}{\partial \tau}; \qquad
\alpha(\tau_f) = n(\tau_f)\int_0^{\tau_f} \frac{b^{\nu}}{n}\frac{\partial u_{\nu}}{\partial \tau} \mathrm{d}\tau + \alpha(0)\,, 
\label{aleq}
\eq
where the above requires one integration per flow line.

Thus the freedom in $\alpha$ reduces to a choice of $\alpha$ on the initial surface $\Omega$, its value at any later time being found by solving the Cauchy problem.  Furthermore, even this initial step may be rendered trivial.  While discussing the condition \eqref{ConditionOnAlpha}, we pointed out that one can add to $\alpha$ any quantity $\Delta \alpha$ obeying $\partial_{\mu}(\Delta \alpha \; u^{\mu}) = 0$.  Reiterating the argument that led to \eqref{aleq}, we find this becomes
\begin{equation*}
\frac{\partial}{\partial \tau}\left(\frac{\Delta \alpha}{n}\right) = 0 \, ,
\end{equation*}
which says that we may choose $\Delta \alpha$ freely on $\Omega$, and the ratio of it over the number density will remain constant along flow lines.  Given this freedom, why not simply pick $\Delta \alpha = -\alpha$ on the initial surface?  So the new $\alpha$ is zero on $\Omega$, the initial conditions are simply $(m^{\mu},h^{\mu}) = ((\rho+p+|b^2|)u^{\mu},b^{\mu})$, and $\alpha$ develops along flow lines according to \eqref{aleq}.  Said integral never actually has to be evaluated, for if one solves the Cauchy problem for $m^{\mu}$ and $h^{\mu}$ (whose equations of motion incorporate the condition $\partial_{\mu}h^{\mu} = 0$), one can then calculate $\alpha$ via \eqref{albe}.  Nonetheless, \eqref{aleq} may be useful as a consistency check on calculations or simulations.  Similarly, the two constraints $u^{\mu}b_{\mu} = 0$ and $u^{\mu}u_{\mu} = 1$ propagate along the flow lines and do not need to be enforced explicitly provided they are true on $\Omega$ initially, though they too remain useful as consistency checks.


We close this discussion by considering a  point that may cause confusion.  Given $(m^\mu, h^\mu)$ on $\Om$ we can certainly calculate $\nabla\cdot\mathbf{h}$,  and $\p h^0/\p \tau$ will be determined by the equations of motion for $(m^\mu, h^\mu)$. Thus, one may wonder how we are free to chose $\al$ and $\p \al /\p \tau$ to make $\p_\mu h^\mu=0$.   The answer lies in the fact that the $(m^\mu, h^\mu)$ system has a solution space that includes  solutions that are not relativistic MHD solutions, and our procedure for picking the quantity selects out those that do indeed correspond --  for these the two ways of determining $\p_\mu h^\mu$ are equivalent. 

\section{Summary}
\label{sec:summary}

We have successfully cast relativistic MHD into a covariant action formalism using a noncanonical bracket.  Along the way, we had to develop a few new ideas with possible consequences beyond our current domain: a modified enthalpy density containing a magnetic ``pressure", a canonical momentum differing from the kinetic momentum by a magnetic term, and a divergenceless magnetic 4-vector possessing a new degeneracy and symmetry.  We presented several closely related additional brackets, and carefully investigated the noninvertibility of the transformations between our original Eulerian quantities and their conjugate momenta.  Many consequences of our formalism were investigated, but many more remain to be covered: for instance, 3+1 reductions, additional Casimirs, the relation to Lagrangian action principles, brackets in systems possessing extra symmetry (e.g. spherical or toroidal), applications to the Godunov numerical scheme in relativity, and conserved quantities related to the $\alpha$ symmetry.  It may be objected that we have, as yet, produced no practical application for our formalism, though it certainly does possess a certain beauty of its own.  However, while practicality usually precedes beauty in physics, the opposite is sometimes the case, reason enough not to disregard that beauty.

\section*{Acknowledgments}
\noindent   FP would like to dedicate this article to his erstwhile  university classmate,  Marcello Anile, who died prematurely in 2007.  ECD and PJM would like to acknowledge helpful conversations with Larry Shepley.  In addition, ECD would like to acknowledge helpful conversations with Wolfgang Rindler, and PJM would like to acknowledge helpful conversations with Zensho Yoshida and Yohei Kawazura.   George Miloshevich also helpfully pointed out an error in an early draft.  ECD and PJM were supported by U.S. Dept.\ of Energy Contract \# DE-FG02-04ER54742.

\appendix
\section{Direct proof of the Jacobi identity}
\label{appendix}

The brackets of \eqref{RelativisticBracket} and \eqref{AlternativeRelativisticBracket} are direct generalizations of the Lie-Poisson form given in Refs.~\onlinecite{morgreene,morgreene82,morrison82} for nonrelativistic MHD, so the Jacobi identity follows from general Lie algebraic and functional derivative properties  (see e.g., Refs.~\onlinecite{morrison82, mm84,marsden,morrison98}).  However, since these will not be known to most readers we include a direct proof in this appendix. 

The Jacobi identity is
\begin{equation} \label{JacobiIdentity}
\{\{F,G\},H\} + \{\{G,H\},F\} + \{\{H,F\},G\} = 0
\end{equation}
for the two brackets \eqref{RelativisticBracket} and \eqref{AlternativeRelativisticBracket}.  

When expanding the expression \eqref{JacobiIdentity}, many terms will contain second functional derivatives, for instance
\begin{equation*}
n h^{\lambda}\frac{\delta G}{\delta m_{\nu}}\left(\partial_{\nu}
\frac{\delta^2 F}{\delta h^{\lambda} \delta m_{\mu}}\right)\partial_{\mu}\frac{\delta H}{\delta n}
\end{equation*}
Thankfully, by a theorem in Ref.~\onlinecite{morrison82}, all such terms cancel for any antisymmetric bracket.  Thus we only have to worry about those terms containing only first functional derivatives.  Starting with the bracket \eqref{RelativisticBracket}, the needed terms are thus
\begin{equation} \label{FunctionalDerivativesofBracket} \begin{aligned}
\frac{\delta \{F,G\}}{\delta n} =& \frac{\delta F}{\delta m_{\mu}}\partial_{\mu}\frac{\delta G}{\delta n} -
\frac{\delta G}{\delta m_{\mu}}\partial_{\mu}\frac{\delta F}{\delta n} + \dots \\
\frac{\delta \{F,G\}}{\delta \sigma} = & \frac{\delta F}{\delta m_{\mu}}\partial_{\mu}\frac{\delta G}{\delta \sigma} - 
\frac{\delta G}{\delta m_{\mu}}\partial_{\mu}\frac{\delta F}{\delta \sigma} + \dots \\
\frac{\delta \{F,G\}}{\delta m_{\mu}} = & \frac{\delta F}{\delta m_{\nu}}\partial_{\nu}\frac{\delta G}{\delta m_{\mu}} - 
\frac{\delta G}{\delta m_{\nu}}\partial_{\nu}\frac{\delta F}{\delta m_{\mu}} + \dots \\
\frac{\delta \{F,G\}}{\delta h^{\mu}} = & \frac{\delta F}{\delta m_{\nu}}\partial_{\nu}\frac{\delta G}{\delta h^{\mu}} - 
\frac{\delta G}{\delta m_{\nu}}\partial_{\nu}\frac{\delta F}{\delta h^{\mu}} \enspace + \\
& \partial_{\mu}\frac{\delta F}{\delta m_{\nu}}\frac{\delta G}{\delta h^{\nu}} -
\partial_{\mu}\frac{\delta G}{\delta m_{\nu}}\frac{\delta F}{\delta h^{\nu}} + \dots
\end{aligned} \end{equation}
with similar expressions for the other two permutations of $F$, $G$, and $H$.  Beginning with this expression, it is to be understood that, in the absence of parentheses, the gradient operators act only on the term immediately to their right; when they are followed by an expression in parentheses, they act as normal.  This convention will remove many superfluous symbols.  The ellipses at the end of each line indicate the terms that may be disregarded thanks to the aforementioned theorem.  Upon inserting the expressions \eqref{FunctionalDerivativesofBracket} into the Jacobi identity \eqref{JacobiIdentity}, all pertinent terms will be linear in the field variables.  Each of these four sets of terms (one for each field variable) must vanish separately.  

The terms linear in $n$ are:
\begin{equation} \label{NPortionofJacobi}
\int\!d^4 x \,  n \left[\left(\frac{\delta F}{\delta m_{\nu}}\partial_{\nu}\frac{\delta G}{\delta m_{\mu}} -
\frac{\delta G}{\delta m_{\nu}}\partial_{\nu}\frac{\delta F}{\delta m_{\mu}}\right)\partial_{\mu}\frac{\delta H}{\delta n} 
- \frac{\delta H}{\delta m_{\mu}}\partial_{\mu}\left(\frac{\delta F}{\delta m_{\nu}}\partial_{\nu}\frac{\delta G}{\delta n}
- \frac{\delta G}{\delta m_{\nu}}\partial_{\nu}\frac{\delta F}{\delta n}\right) + \underset{F,G,H}{\mathlarger{\circlearrowright}}
\right]  
\end{equation}
where the circle symbol indicates permutation in $F$, $G$, and $H$.  Inside the square braces, the collected second derivative terms are
\begin{equation*} \begin{aligned}
-\frac{\delta H}{\delta m_{\mu}}\frac{\delta F}{\delta m_{\nu}}\partial^2_{\mu\nu}\frac{\delta G}{\delta n}
+\frac{\delta H}{\delta m_{\mu}}\frac{\delta G}{\delta m_{\nu}}\partial^2_{\mu\nu}\frac{\delta F}{\delta n}
-\frac{\delta F}{\delta m_{\mu}}\frac{\delta G}{\delta m_{\nu}}\partial^2_{\mu\nu}\frac{\delta H}{\delta n} \\
+\frac{\delta F}{\delta m_{\mu}}\frac{\delta H}{\delta m_{\nu}}\partial^2_{\mu\nu}\frac{\delta G}{\delta n}
-\frac{\delta G}{\delta m_{\mu}}\frac{\delta H}{\delta m_{\nu}}\partial^2_{\mu\nu}\frac{\delta F}{\delta n}
+\frac{\delta G}{\delta m_{\mu}}\frac{\delta F}{\delta m_{\nu}}\partial^2_{\mu\nu}\frac{\delta H}{\delta n}
\end{aligned} \end{equation*}
which vanish due to the fact that second (partial) derivatives commute.  The remaining terms linear in $n$, keeping the same order they have in the Jacobi identity, follow:
\begin{equation*} \begin{aligned}
\frac{\delta F}{\delta m_{\nu}}\partial_{\nu}\frac{\delta G}{\delta m_{\mu}}\partial_{\mu}\frac{\delta H}{\delta n}^{\textcircled{2}} -&
\frac{\delta G}{\delta m_{\nu}}\partial_{\nu}\frac{\delta F}{\delta m_{\mu}}\partial_{\mu}\frac{\delta H}{\delta n}^{\textcircled{6}} -
\frac{\delta H}{\delta m_{\mu}}\partial_{\mu}\frac{\delta F}{\delta m_{\nu}}\partial_{\nu}\frac{\delta G}{\delta n}^{\textcircled{3}} +
\frac{\delta H}{\delta m_{\mu}}\partial_{\mu}\frac{\delta G}{\delta m_{\nu}}\partial_{\nu}\frac{\delta F}{\delta n}^{\textcircled{1}} + \\
\frac{\delta G}{\delta m_{\nu}}\partial_{\nu}\frac{\delta H}{\delta m_{\mu}}\partial_{\mu}\frac{\delta F}{\delta n}^{\textcircled{5}} -&
\frac{\delta H}{\delta m_{\nu}}\partial_{\nu}\frac{\delta G}{\delta m_{\mu}}\partial_{\mu}\frac{\delta F}{\delta n}^{\textcircled{1}} -
\frac{\delta F}{\delta m_{\mu}}\partial_{\mu}\frac{\delta G}{\delta m_{\nu}}\partial_{\nu}\frac{\delta H}{\delta n}^{\textcircled{2}} +
\frac{\delta F}{\delta m_{\mu}}\partial_{\mu}\frac{\delta H}{\delta m_{\nu}}\partial_{\nu}\frac{\delta G}{\delta n}^{\textcircled{4}} + \\
\frac{\delta H}{\delta m_{\nu}}\partial_{\nu}\frac{\delta F}{\delta m_{\mu}}\partial_{\mu}\frac{\delta G}{\delta n}^{\textcircled{3}} -&
\frac{\delta F}{\delta m_{\nu}}\partial_{\nu}\frac{\delta H}{\delta m_{\mu}}\partial_{\mu}\frac{\delta G}{\delta n}^{\textcircled{4}} -
\frac{\delta G}{\delta m_{\mu}}\partial_{\mu}\frac{\delta H}{\delta m_{\nu}}\partial_{\nu}\frac{\delta F}{\delta n}^{\textcircled{5}} +
\frac{\delta G}{\delta m_{\mu}}\partial_{\mu}\frac{\delta F}{\delta m_{\nu}}\partial_{\nu}\frac{\delta H}{\delta n}^{\textcircled{6}}
\end{aligned} \end{equation*}
They vanish in pairs, as labeled by the circled numbers.

So all the terms linear in $n$ have vanished from the Jacobi identity.  However, the terms linear in $\sigma$ are identical, but with functional derivatives $\delta /\delta n$ replaced by $\delta / \delta \sigma$.  So the $\sigma$ terms vanish by an identical calculation.  Moreover, the $m_{\lambda}$ terms do as well: the $\delta / \delta n$ are replaced with $\delta / \delta m_{\lambda}$, contracted with the remaining $m_{\lambda}$ term outside the square brackets of its version of \eqref{NPortionofJacobi}, and the calculation proceeds as before.

The only terms remaining to be checked are those linear in $h^{\lambda}$; unfortunately, there are quite a few:
\begin{equation*} \begin{aligned}
\int\!d^4 x\, h^{\lambda} \left[\left(\frac{\delta F}{\delta m_{\mu}}\partial_{\mu}\frac{\delta G}{\delta m_{\nu}}
 - \frac{\delta G}{\delta m_{\mu}}\partial_{\mu}\frac{\delta F}{\delta m_{\nu}}\right)\partial_{\nu}
\frac{\delta H}{\delta h^{\lambda}}^{\textcircled{1}} \right.  & \\
- \frac{\delta H}{\delta m_{\nu}}\partial_{\nu}\left(\frac{\delta F}{\delta m_{\mu}}\partial_{\mu}\frac{\delta G}{\delta h^{\lambda}}
-\frac{\delta G}{\delta m_{\mu}}\partial_{\mu}\frac{\partial F}{\partial h^{\lambda}}\right)^{\textcircled{1}} 
& - \frac{\delta H}{\delta m_{\nu}}\partial_{\nu}\left(\partial_{\lambda}\frac{\delta F}{\delta m_{\mu}}\frac{\delta G}{\delta h^{\mu}}
- \partial_{\lambda}\frac{\delta G}{\delta m_{\mu}}\frac{\delta F}{\delta h^{\mu}}\right) \\
+ \partial_{\lambda}\left(\frac{\delta F}{\delta m_{\mu}}\partial_{\mu}\frac{\delta G}{\delta m_{\nu}}
- \frac{\delta G}{\delta m_{\mu}}\partial_{\mu}\frac{\delta F}{\delta m_{\nu}}\right)\frac{\delta H}{\delta h^{\nu}} \\
- \partial_{\lambda}\frac{\delta H}{\delta m_{\nu}}\left(\frac{\delta F}{\delta m_{\mu}}\partial_{\mu}\frac{\delta G}{\delta h^{\nu}}
- \frac{\delta G}{\delta m_{\mu}}\partial_{\mu}\frac{\delta F}{\delta h^{\nu}} \right. & \left. \left.
+ \partial_{\nu}\frac{\delta F}{\delta m_{\mu}}\frac{\delta G}{\delta h^{\mu}}
- \partial_{\nu}\frac{\delta G}{\delta m_{\mu}}\frac{\delta F}{\delta h^{\mu}}\right)
+ \underset{F,G,H}{\mathlarger{\circlearrowright}} \right] 
\end{aligned} \end{equation*}
The terms labelled by a circled ``one" produce a calculation identical to that already performed, and thus cancel.  From the remaining terms, we first gather all the second derivative ones inside the square braces:
\begin{equation*} \begin{aligned}
- & \frac{\delta H}{\delta m_{\nu}}\frac{\delta G}{\delta h^{\mu}}\partial^2_{\lambda\nu}\frac{\delta F}{\delta m_{\mu}}^{\textcircled{5}}
+ \frac{\delta H}{\delta m_{\nu}}\frac{\delta F}{\delta h^{\mu}}\partial^2_{\lambda\nu}\frac{\delta G}{\delta m_{\mu}}^{\textcircled{2}}
+ \frac{\delta F}{\delta m_{\mu}}\frac{\delta H}{\delta h^{\nu}}\partial^2_{\lambda\mu}\frac{\delta G}{\delta m_{\nu}}^{\textcircled{1}}
- \frac{\delta G}{\delta m_{\mu}}\frac{\delta H}{\delta h^{\nu}}\partial^2_{\lambda\mu}\frac{\delta F}{\delta m_{\nu}}^{\textcircled{4}} \\
- & \frac{\delta F}{\delta m_{\nu}}\frac{\delta H}{\delta h^{\mu}}\partial^2_{\lambda\nu}\frac{\delta G}{\delta m_{\mu}}^{\textcircled{1}}
+ \frac{\delta F}{\delta m_{\nu}}\frac{\delta G}{\delta h^{\mu}}\partial^2_{\lambda\nu}\frac{\delta H}{\delta m_{\mu}}^{\textcircled{6}}
+ \frac{\delta G}{\delta m_{\mu}}\frac{\delta F}{\delta h^{\nu}}\partial^2_{\lambda\mu}\frac{\delta H}{\delta m_{\nu}}^{\textcircled{3}}
- \frac{\delta H}{\delta m_{\mu}}\frac{\delta F}{\delta h^{\nu}}\partial^2_{\lambda\mu}\frac{\delta G}{\delta m_{\nu}}^{\textcircled{2}} \\
- & \frac{\delta G}{\delta m_{\nu}}\frac{\delta F}{\delta h^{\mu}}\partial^2_{\lambda\nu}\frac{\delta H}{\delta m_{\mu}}^{\textcircled{3}}
+ \frac{\delta G}{\delta m_{\nu}}\frac{\delta H}{\delta h^{\mu}}\partial^2_{\lambda\nu}\frac{\delta F}{\delta m_{\mu}}^{\textcircled{4}}
+ \frac{\delta H}{\delta m_{\mu}}\frac{\delta G}{\delta h^{\nu}}\partial^2_{\lambda\mu}\frac{\delta F}{\delta m_{\nu}}^{\textcircled{5}}
- \frac{\delta F}{\delta m_{\mu}}\frac{\delta G}{\delta h^{\nu}}\partial^2_{\lambda\mu}\frac{\delta H}{\delta m_{\nu}}^{\textcircled{6}} \\
\end{aligned} \end{equation*}
They cancel in pairs.  Finally, the remaining terms, in the same order and bearing the same indices as in the Jacobi identity, are:
\begin{equation*} \begin{aligned}
- & \frac{\delta H}{\delta m_{\nu}}\partial_{\lambda}\frac{\delta F}{\delta m_{\mu}}\partial_{\nu}\frac{\delta G}{\delta h^{\mu}}
^{\textcircled{3}} 
+ \frac{\delta H}{\delta m_{\nu}}\partial_{\lambda}\frac{\delta G}{\delta m_{\mu}}\partial_{\nu}\frac{\delta F}{\delta h^{\mu}}
^{\textcircled{9}} 
+ \frac{\delta H}{\delta h^{\nu}}\partial_{\lambda}\frac{\delta F}{\delta m_{\mu}}\partial_{\mu}\frac{\delta G}{\delta m_{\nu}}
^{\textcircled{4}}
- \frac{\delta H}{\delta h^{\nu}}\partial_{\lambda}\frac{\delta G}{\delta m_{\mu}}\partial_{\mu}\frac{\delta F}{\delta m_{\nu}}
^{\textcircled{12}} \\
- & \frac{\delta F}{\delta m_{\mu}}\partial_{\lambda}\frac{\delta H}{\delta m_{\nu}}\partial_{\mu}\frac{\delta G}{\delta h^{\nu}}
^{\textcircled{1}}
+ \frac{\delta G}{\delta m_{\mu}}\partial_{\lambda}\frac{\delta H}{\delta m_{\nu}}\partial_{\mu}\frac{\delta F}{\delta h^{\nu}}
^{\textcircled{5}}
- \frac{\delta G}{\delta h^{\mu}}\partial_{\lambda}\frac{\delta H}{\delta m_{\nu}}\partial_{\nu}\frac{\delta F}{\delta m_{\mu}}
^{\textcircled{7}}
+ \frac{\delta F}{\delta h^{\mu}}\partial_{\lambda}\frac{\delta H}{\delta m_{\nu}}\partial_{\nu}\frac{\delta G}{\delta m_{\mu}}
^{\textcircled{2}} \\
- & \frac{\delta F}{\delta m_{\nu}}\partial_{\lambda}\frac{\delta G}{\delta m_{\mu}}\partial_{\nu}\frac{\delta H}{\delta h^{\mu}}
^{\textcircled{10}} 
+ \frac{\delta F}{\delta m_{\nu}}\partial_{\lambda}\frac{\delta H}{\delta m_{\mu}}\partial_{\nu}\frac{\delta G}{\delta h^{\mu}}
^{\textcircled{1}} 
+ \frac{\delta F}{\delta h^{\nu}}\partial_{\lambda}\frac{\delta G}{\delta m_{\mu}}\partial_{\mu}\frac{\delta H}{\delta m_{\nu}}
^{\textcircled{11}}
- \frac{\delta F}{\delta h^{\nu}}\partial_{\lambda}\frac{\delta H}{\delta m_{\mu}}\partial_{\mu}\frac{\delta G}{\delta m_{\nu}}
^{\textcircled{2}} \\
- & \frac{\delta G}{\delta m_{\mu}}\partial_{\lambda}\frac{\delta F}{\delta m_{\nu}}\partial_{\mu}\frac{\delta H}{\delta h^{\nu}}
^{\textcircled{6}}
+ \frac{\delta H}{\delta m_{\mu}}\partial_{\lambda}\frac{\delta F}{\delta m_{\nu}}\partial_{\mu}\frac{\delta G}{\delta h^{\nu}}
^{\textcircled{3}}
- \frac{\delta H}{\delta h^{\mu}}\partial_{\lambda}\frac{\delta F}{\delta m_{\nu}}\partial_{\nu}\frac{\delta G}{\delta m_{\mu}}
^{\textcircled{4}}
+ \frac{\delta G}{\delta h^{\mu}}\partial_{\lambda}\frac{\delta F}{\delta m_{\nu}}\partial_{\nu}\frac{\delta H}{\delta m_{\mu}}
^{\textcircled{8}} \\
- & \frac{\delta G}{\delta m_{\nu}}\partial_{\lambda}\frac{\delta H}{\delta m_{\mu}}\partial_{\nu}\frac{\delta F}{\delta h^{\mu}}
^{\textcircled{5}} 
+ \frac{\delta G}{\delta m_{\nu}}\partial_{\lambda}\frac{\delta F}{\delta m_{\mu}}\partial_{\nu}\frac{\delta H}{\delta h^{\mu}}
^{\textcircled{6}} 
+ \frac{\delta G}{\delta h^{\nu}}\partial_{\lambda}\frac{\delta H}{\delta m_{\mu}}\partial_{\mu}\frac{\delta F}{\delta m_{\nu}}
^{\textcircled{7}}
- \frac{\delta G}{\delta h^{\nu}}\partial_{\lambda}\frac{\delta F}{\delta m_{\mu}}\partial_{\mu}\frac{\delta H}{\delta m_{\nu}}
^{\textcircled{8}} \\
- & \frac{\delta H}{\delta m_{\mu}}\partial_{\lambda}\frac{\delta G}{\delta m_{\nu}}\partial_{\mu}\frac{\delta F}{\delta h^{\nu}}
^{\textcircled{9}}
+ \frac{\delta F}{\delta m_{\mu}}\partial_{\lambda}\frac{\delta G}{\delta m_{\nu}}\partial_{\mu}\frac{\delta H}{\delta h^{\nu}}
^{\textcircled{10}}
- \frac{\delta F}{\delta h^{\mu}}\partial_{\lambda}\frac{\delta G}{\delta m_{\nu}}\partial_{\nu}\frac{\delta H}{\delta m_{\mu}}
^{\textcircled{11}}
+ \frac{\delta H}{\delta h^{\mu}}\partial_{\lambda}\frac{\delta G}{\delta m_{\nu}}\partial_{\nu}\frac{\delta F}{\delta m_{\mu}}
^{\textcircled{12}} \\
\end{aligned} \end{equation*}
They also cancel in pairs, establishing the Jacobi identity.  This derivation is also valid in curved spacetimes, for the functional derivative cancels out a factor of $\sqrt{-g}$, and there is no integration by parts to catch another such factor.

Next we will perform a similar calculation for the alternative bracket \eqref{AlternativeRelativisticBracket}.  While the same kinds of terms appear as above, there is no longer a complete cancellation.  Most of the functional derivatives \eqref{FunctionalDerivativesofBracket} are unchanged, the only differing one being
\begin{equation*}
\frac{\delta\{F,G\}}{\delta h^{\mu}} = \frac{\delta F}{\delta m_{\nu}}\partial_{\nu}\frac{\partial G}{\partial h^{\mu}}
- \frac{\delta G}{\delta m_{\nu}}\partial_{\nu}\frac{\partial F}{\partial h^{\mu}}
+ \partial_{\mu}\frac{\delta F}{\delta h^{\nu}}\frac{\delta G}{\delta m_{\nu}}
- \partial_{\mu}\frac{\delta G}{\delta h^{\nu}}\frac{\delta F}{\delta m_{\nu}} + \dots
\end{equation*}
with the ellipsis again indicating terms with second functional derivatives, all of which can be disregarded.

The terms of the Jacobi identity once more appear in four sets, each linear in one of the field variables.  The $n$, $\sigma$, and $m^{\lambda}$ terms involve no derivatives with respect to $h^{\lambda}$, and are thus unchanged: they cancel as before.  Only the $h^{\lambda}$ terms differ.  They read:
\begin{equation*} \begin{aligned}
\int\!d^4 x\, h^{\lambda}\left[\left(\frac{\delta F}{\delta m_{\nu}}\partial_{\nu}\frac{\delta G}{\delta m_{\mu}} -
\frac{\delta G}{\delta m_{\nu}}\partial_{\nu}\frac{\delta F}{\delta m_{\mu}}\right)
\partial_{\mu}\frac{\delta H}{\delta h^{\lambda}}^{\textcircled{1}} \right. \\
- \frac{\delta H}{\delta m_{\nu}}\partial_{\nu}\left(\frac{\delta F}{\delta m_{\mu}}\partial_{\mu}\frac{\delta G}{\delta h^{\lambda}}
- \frac{\delta G}{\delta m_{\mu}}\partial_{\mu}\frac{\delta F}{\delta h^{\lambda}}\right)^{\textcircled{1}}
- \frac{\delta H}{\delta m_{\nu}}\partial_{\nu}\left(\partial_{\lambda}\frac{\delta F}{\delta h^{\mu}}\frac{\delta G}{\delta m_{\mu}}
- \partial_{\lambda}\frac{\delta G}{\delta h^{\mu}}\frac{\delta F}{\delta m_{\mu}}\right) \\
+ \partial_{\lambda}\left(\frac{\delta F}{\delta m_{\mu}}\partial_{\mu}\frac{\delta G}{\delta h^{\nu}} -
\frac{\delta G}{\delta m_{\mu}}\partial_{\mu}\frac{\delta F}{\delta h^{\nu}}
+ \partial_{\nu}\frac{\delta F}{\delta h^{\mu}}\frac{\delta G}{\delta m_{\mu}}
- \partial_{\nu}\frac{\delta G}{\delta h^{\mu}}\frac{\delta F}{\delta m_{\mu}}\right)\frac{\delta H}{\delta m_{\nu}} \\
- \left. \partial_{\lambda}\frac{\delta H}{\delta h^{\nu}}\left(\frac{\delta F}{\delta m_{\mu}}\partial_{\mu}\frac{\delta G}{\delta m_{\nu}}
- \frac{\delta G}{\delta m_{\mu}}\partial_{\mu}\frac{\delta F}{\delta m_{\nu}}\right)
+ \underset{F,G,H}{\mathlarger{\circlearrowright}} \right]  
\end{aligned} \end{equation*}
The terms labelled with a circled ``one" cancel as in the previous bracket.  The collected second derivative terms are
\begin{equation*} \begin{aligned}
- & \frac{\delta H}{\delta m_{\nu}}\frac{\delta G}{\delta m_{\mu}}\partial^2_{\nu\lambda}\frac{\delta F}{\delta h^{\mu}}^{\textcircled{2}}
+ \frac{\delta H}{\delta m_{\nu}}\frac{\delta F}{\delta m_{\mu}}\partial^2_{\nu\lambda}\frac{\delta G}{\delta h^{\mu}}^{\textcircled{1}}
+ \frac{\delta F}{\delta m_{\mu}}\frac{\delta H}{\delta m_{\nu}}\partial^2_{\lambda\mu}\frac{\delta G}{\delta h^{\nu}} \\
- & \frac{\delta G}{\delta m_{\mu}}\frac{\delta H}{\delta m_{\nu}}\partial^2_{\lambda\mu}\frac{\delta F}{\delta h^{\nu}}
+ \frac{\delta H}{\delta m_{\nu}}\frac{\delta G}{\delta m_{\mu}}\partial^2_{\lambda\nu}\frac{\delta F}{\delta h^{\mu}}^{\textcircled{2}}
- \frac{\delta H}{\delta m_{\nu}}\frac{\delta F}{\delta m_{\mu}}\partial^2_{\lambda\nu}\frac{\delta G}{\delta h^{\mu}}^{\textcircled{1}}
+ \underset{F,G,H}{\mathlarger{\circlearrowright}} \\
= & \frac{\delta F}{\delta m_{\mu}}\frac{\delta H}{\delta m_{\nu}}\partial^2_{\lambda\mu}\frac{\delta G}{\delta h^{\nu}}
- \frac{\delta G}{\delta m_{\mu}}\frac{\delta H}{\delta m_{\nu}}\partial^2_{\lambda\mu}\frac{\delta F}{\delta h^{\nu}}
+ \frac{\delta G}{\delta m_{\mu}}\frac{\delta F}{\delta m_{\nu}}\partial^2_{\lambda\mu}\frac{\delta H}{\delta h^{\nu}} \\
- & \frac{\delta H}{\delta m_{\mu}}\frac{\delta F}{\delta m_{\nu}}\partial^2_{\lambda\mu}\frac{\delta G}{\delta h^{\nu}}
+ \frac{\delta H}{\delta m_{\mu}}\frac{\delta G}{\delta m_{\nu}}\partial^2_{\lambda\mu}\frac{\delta F}{\delta h^{\nu}} 
-  \frac{\delta F}{\delta m_{\mu}}\frac{\delta G}{\delta m_{\nu}}\partial^2_{\lambda\mu}\frac{\delta H}{\delta h^{\nu}}
\end{aligned} \end{equation*}
Six terms do not cancel.  The other terms (i.e. those that are not second derivatives) are
\begin{equation*} \begin{aligned}
- & \frac{\delta H}{\delta m_{\nu}}\partial_{\lambda}\frac{\delta F}{\delta h^{\mu}}\partial_{\nu}\frac{\delta G}{\delta m_{\mu}}
^{\textcircled{2}}
+  \frac{\delta H}{\delta m_{\nu}}\partial_{\lambda}\frac{\delta G}{\delta h^{\mu}}\partial_{\nu}\frac{\delta F}{\delta m_{\mu}}
^{\textcircled{5}}
+  \frac{\delta H}{\delta m_{\nu}}\partial_{\lambda}\frac{\delta F}{\delta m_{\mu}}\partial_{\mu}\frac{\delta G}{\delta h^{\nu}}
-  \frac{\delta H}{\delta m_{\nu}}\partial_{\lambda}\frac{\delta G}{\delta m_{\mu}}\partial_{\mu}\frac{\delta F}{\delta h^{\nu}} \\
+ & \frac{\delta H}{\delta m_{\nu}}\partial_{\lambda}\frac{\delta G}{\delta m_{\mu}}\partial_{\nu}\frac{\delta F}{\delta h^{\mu}}
-  \frac{\delta H}{\delta m_{\nu}}\partial_{\lambda}\frac{\delta F}{\delta m_{\mu}}\partial_{\nu}\frac{\delta G}{\delta h^{\mu}}
-  \frac{\delta F}{\delta m_{\mu}}\partial_{\lambda}\frac{\delta H}{\delta h^{\nu}}\partial_{\mu}\frac{\delta G}{\delta m_{\nu}}
^{\textcircled{1}}
+  \frac{\delta G}{\delta m_{\mu}}\partial_{\lambda}\frac{\delta H}{\delta h^{\nu}}\partial_{\mu}\frac{\delta F}{\delta m_{\nu}}
^{\textcircled{3}} \\
- & \frac{\delta F}{\delta m_{\nu}}\partial_{\lambda}\frac{\delta G}{\delta h^{\mu}}\partial_{\nu}\frac{\delta H}{\delta m_{\mu}}
^{\textcircled{6}}
+  \frac{\delta F}{\delta m_{\nu}}\partial_{\lambda}\frac{\delta H}{\delta h^{\mu}}\partial_{\nu}\frac{\delta G}{\delta m_{\mu}}
^{\textcircled{1}}
+  \frac{\delta F}{\delta m_{\nu}}\partial_{\lambda}\frac{\delta G}{\delta m_{\mu}}\partial_{\mu}\frac{\delta H}{\delta h^{\nu}}
-  \frac{\delta F}{\delta m_{\nu}}\partial_{\lambda}\frac{\delta H}{\delta m_{\mu}}\partial_{\mu}\frac{\delta G}{\delta h^{\nu}} \\
+ & \frac{\delta F}{\delta m_{\nu}}\partial_{\lambda}\frac{\delta H}{\delta m_{\mu}}\partial_{\nu}\frac{\delta G}{\delta h^{\mu}}
-  \frac{\delta F}{\delta m_{\nu}}\partial_{\lambda}\frac{\delta G}{\delta m_{\mu}}\partial_{\nu}\frac{\delta H}{\delta h^{\mu}}
-  \frac{\delta G}{\delta m_{\mu}}\partial_{\lambda}\frac{\delta F}{\delta h^{\nu}}\partial_{\mu}\frac{\delta H}{\delta m_{\nu}}
^{\textcircled{4}}
+  \frac{\delta H}{\delta m_{\mu}}\partial_{\lambda}\frac{\delta F}{\delta h^{\nu}}\partial_{\mu}\frac{\delta G}{\delta m_{\nu}}
^{\textcircled{2}} \\
- & \frac{\delta G}{\delta m_{\nu}}\partial_{\lambda}\frac{\delta H}{\delta h^{\mu}}\partial_{\nu}\frac{\delta F}{\delta m_{\mu}}
^{\textcircled{3}}
+  \frac{\delta G}{\delta m_{\nu}}\partial_{\lambda}\frac{\delta F}{\delta h^{\mu}}\partial_{\nu}\frac{\delta H}{\delta m_{\mu}}
^{\textcircled{4}}
+  \frac{\delta G}{\delta m_{\nu}}\partial_{\lambda}\frac{\delta H}{\delta m_{\mu}}\partial_{\mu}\frac{\delta F}{\delta h^{\nu}}
-  \frac{\delta G}{\delta m_{\nu}}\partial_{\lambda}\frac{\delta F}{\delta m_{\mu}}\partial_{\mu}\frac{\delta H}{\delta h^{\nu}} \\
+ & \frac{\delta G}{\delta m_{\nu}}\partial_{\lambda}\frac{\delta F}{\delta m_{\mu}}\partial_{\nu}\frac{\delta H}{\delta h^{\mu}}
-  \frac{\delta G}{\delta m_{\nu}}\partial_{\lambda}\frac{\delta H}{\delta m_{\mu}}\partial_{\nu}\frac{\delta F}{\delta h^{\mu}}
-  \frac{\delta H}{\delta m_{\mu}}\partial_{\lambda}\frac{\delta G}{\delta h^{\nu}}\partial_{\mu}\frac{\delta F}{\delta m_{\nu}}
^{\textcircled{5}}
+  \frac{\delta F}{\delta m_{\mu}}\partial_{\lambda}\frac{\delta G}{\delta h^{\nu}}\partial_{\mu}\frac{\delta H}{\delta m_{\nu}}
^{\textcircled{6}} \\
\end{aligned} \end{equation*}
This time twelve terms do not cancel.  All told, eighteen terms remain, which collect in groups of three.  Each group reduces to a gradient with a $\partial_{\lambda}$ pulled outside the expression.  The whole Jacobi identity simplifies to
\begin{equation*} \begin{aligned}
\{\{F,G\},H\} + \{\{G,H\},F\} + \{\{H,F\},G\} \\
= \int\!d^4 x\,  h^{\lambda}\partial_{\lambda}\left(
\frac{\delta F}{\delta m_{\nu}}\frac{\delta G}{\delta m_{\mu}}\partial_{\mu}\frac{\delta H}{\delta h^{\nu}} -
\frac{\delta G}{\delta m_{\nu}}\frac{\delta F}{\delta m_{\mu}}\partial_{\mu}\frac{\delta H}{\delta h^{\nu}} +
\frac{\delta G}{\delta m_{\nu}}\frac{\delta H}{\delta m_{\mu}}\partial_{\mu}\frac{\delta F}{\delta h^{\nu}} - \right. \\
\left. \frac{\delta H}{\delta m_{\nu}}\frac{\delta G}{\delta m_{\mu}}\partial_{\mu}\frac{\delta F}{\delta h^{\nu}} +
\frac{\delta H}{\delta m_{\nu}}\frac{\delta F}{\delta m_{\mu}}\partial_{\mu}\frac{\delta G}{\delta h^{\nu}} - 
\frac{\delta F}{\delta m_{\nu}}\frac{\delta H}{\delta m_{\mu}}\partial_{\mu}\frac{\delta G}{\delta h^{\nu}} \right)  
\end{aligned}\end{equation*}

An integration by parts shows that the Jacobi identity is satisfied if $h^{\nu}_{\enspace ,\nu} = 0$.  In a curved spacetime, the above expression is the same, except that $d^4x$ becomes $\sqrt{-g} d^4x$.  The integration by parts catches this extra factor, yielding $(h^{\nu} \sqrt{-g})_{, \nu} = h^{\nu}_{\enspace ; \nu} = 0$ as a requirement for the Jacobi identity.

\bibliography{relMHD}

\end{document}